\documentclass[fleqn,12pt]{wlscirep}
\usepackage[utf8]{inputenc}
\usepackage[T1]{fontenc}
\usepackage[numbers,super]{natbib}
\usepackage{caption}

\newcommand{\araa}{Annu. Rev. Astron. Astrophys.}   
\newcommand{\aj}{Astron. J.}   
\newcommand{\apj}{Astrophys. J.}   
\newcommand{\apjl}{Astrophys. J. Lett.}   
\newcommand{\apjs}{Astrophys. J. Suppl. Ser.}   
\newcommand{\aap}{Astron. Astrophys.}   
\newcommand{\mnras}{Mon. Not. R. Astron. Soc.}   

\newcommand{\nat}{Nature} 
\newcommand{\pasj}{Publ. Astron. Soc. Jpn}   
\newcommand{\pasp}{Publ. Astron. Soc. Pacific}   

\newcommand{\arcsec}{\mbox{\ensuremath{^{\prime\prime}}}}
\newcommand{\farcs}{\mbox{\ensuremath{.\!\!^{\prime\prime}}}}

\newcommand{\jw}{{\it JWST}}
\newcommand{\galight}{\textsf{galight}}
\newcommand{\sersic}{S\'ersic}

\usepackage{color,subfigure}

\newcommand{\lt}{\ensuremath <}
\newcommand{\gt}{\ensuremath >}
\makeatother
\newcommand{\targetone}{J2236$+$0032}
\newcommand{\targettwo}{J1512$+$4422}

\defcitealias{Ding23}{D23}

\graphicspath{{./}{figures_final/}}

\title{A Post-Starburst Pathway for the Formation of Massive Galaxies and Black Holes at $\mathbf{z>6}$}

\author[1,2,3,4,5,*]{Masafusa Onoue}
\author[6,1,*]{Xuheng Ding}
\author[1,3,7,8]{John D. Silverman}
\author[9]{Yoshiki Matsuoka}
\author[10,11]{Takuma Izumi}
\author[12]{Michael A. Strauss}
\author[12]{Charlotte Ward}
\author[12]{Camryn L. Phillips}
\author[7,13,14]{Kei Ito}
\author[15,16]{Irham T. Andika}
\author[17]{Kentaro Aoki}
\author[7]{Junya Arita}
\author[18]{Shunsuke Baba}
\author[19]{Rebekka Bieri}
\author[5,20]{Sarah E. I. Bosman}
\author[21]{Anna-Christina Eilers}
\author[22]{Seiji Fujimoto}
\author[5]{Melanie Habouzit}
\author[23,24]{Zoltan Haiman}
\author[10,25]{Masatoshi Imanishi}
\author[2]{Kohei Inayoshi}
\author[26,27]{Kazushi Iwasawa}
\author[4]{Knud Jahnke}
\author[7,28]{Nobunari Kashikawa}
\author[29]{Toshihiro Kawaguchi}
\author[30,28]{Kotaro Kohno}
\author[31]{Chien-Hsiu Lee}
\author[32]{Junyao Li}
\author[33]{Alessandro Lupi}
\author[34]{Jianwei Lyu}
\author[9]{Tohru Nagao}
\author[35]{Roderik Overzier}
\author[36]{Jan-Torge Schindler}
\author[37]{Malte Schramm}
\author[23]{Matthew T. Scoggins}
\author[7,28]{Kazuhiro Shimasaku}
\author[38,10,39]{Yoshiki Toba}
\author[40]{Benny Trakhtenbrot}
\author[41]{Maxime Trebitsch}
\author[42]{Tommaso Treu}
\author[43,44]{Hideki Umehata}
\author[35]{Bram Venemans}
\author[45,34]{Marianne Vestergaard}
\author[46]{Marta Volonteri}
\author[4]{Fabian Walter}
\author[47]{Feige Wang}
\author[47]{Jinyi Yang}
\author[34]{Haowen Zhang}

\affil[1]{Kavli Institute for the Physics and Mathematics of the Universe (WPI),The University of Tokyo Institutes for Advanced Study, The University of Tokyo, Kashiwa, Chiba 277-8583, Japan}
\affil[2]{Kavli Institute for Astronomy and Astrophysics, Peking University, Beijing 100871, China}
\affil[3]{Center for Data-Driven Discovery, Kavli IPMU (WPI), UTIAS, The University of Tokyo, Kashiwa, Chiba 277-8583, Japan}
\affil[4]{Max-Planck-Institut f\"{u}r Astronomie, K\"{o}nigstuhl 17, D-69117 Heidelberg, Germany}
\affil[5]{Waseda Institute for Advanced Study (WIAS), Waseda University, 1-21-1, Nishi-Waseda, Shinjuku, Tokyo 169-0051, Japan}
\affil[6]{School of Physics and Technology, Wuhan University, Wuhan 430072, China}
\affil[7]{Department of Astronomy, School of Science, The University of Tokyo, 7-3-1 Hongo, Bunkyo, Tokyo 113-0033, Japan}
\affil[8]{Center for Astrophysical Sciences, Department of Physics \& Astronomy, Johns Hopkins University, Baltimore, MD 21218, USA}
\affil[9]{Research Center for Space and Cosmic Evolution, Ehime University, 2-5 Bunkyo-cho, Matsuyama, Ehime 790-8577, Japan}
\affil[10]{National Astronomical Observatory of Japan, Osawa, Mitaka, Tokyo 181-8588, Japan}
\affil[11]{Department of Physics, Graduate School of Science, Tokyo Metropolitan University, 1-1 Minami-Osawa, Hachioji, Tokyo 192-0397, Japan}
\affil[12]{Department of Astrophysical Sciences, Princeton University, 4 Ivy Lane, Princeton, NJ 08544, USA}
\affil[13]{Cosmic Dawn Center (DAWN), Copenhagen, Denmark}
\affil[14]{DTU Space, Technical University of Denmark, Elektrovej 327, DK-2800 Kongens Lyngby, Denmark}
\affil[15]{Technical University of Munich, TUM School of Natural Sciences, Department of Physics, James-Franck-Str. 1, D-85748 Garching, Germany}
\affil[16]{Max-Planck-Institut f\"{u}r Astrophysik, Karl-Schwarzschild-Str. 1, D-85748 Garching, Germany}
\affil[17]{Subaru Telescope, National Astronomical Observatory of Japan, 650 North A’ohoku Pace, Hilo, Hawaii 96720, USA}
\affil[18]{Institute of Space and Astronautical Science, Japan Aerospace Exploration Agency, 3-1-1 Yoshinodai, Chuo-ku, Sagamihara, Kanagawa 252-5210, Japan}
\affil[19]{Department of Astrophysics, University of Zurich, Zurich, Switzerland}
\affil[20]{Institute for Theoretical Physics, Heidelberg University, Philosophenweg 12, D-69120, Heidelberg, Germany}
\affil[21]{MIT Kavli Institute for Astrophysics and Space Research, 77 Massachusetts Avenue, Cambridge, 02139, Massachusetts, USA}
\affil[22]{Department of Astronomy, The University of Texas at Austin, Austin, TX 78712, USA}
\affil[23]{Department of Astronomy, Columbia University, New York, NY 10027, USA}
\affil[24]{Department of Physics, Columbia University, New York, NY 10027, USA}
\affil[25]{Department of Astronomy, School of Science, Graduate University for Advanced Studies (SOKENDAI), Mitaka, Tokyo 181-8588, Japan}
\affil[26]{Institut de Ci\`encies del Cosmos (ICCUB), Universitat de Barcelona (IEEC-UB), Mart\'i i Franqu\`es, 1, 08028 Barcelona, Spain}
\affil[27]{ICREA, Pg. Llu\'is Companys 23, 08010 Barcelona, Spain}
\affil[28]{Research Center for the Early Universe, The University of Tokyo, 7-3-1 Hongo, Bunkyo-ku, Tokyo 113-0033, Japan}
\affil[29]{Graduate School of Science and Engineering, University of Toyama, Gofuku 3190, Toyama 930-8555, Japan}
\affil[30]{Institute of Astronomy, Graduate School of Science, The University of Tokyo, 2-21-1 Osawa, Mitaka, Tokyo 181-0015, Japan}
\affil[31]{W. M. Keck Observatory, 65-1120 Mamalahoa Hwy, Kamuela, HI 96743, USA}
\affil[32]{Department of Astronomy, University of Illinois at Urbana-Champaign, Urbana, IL, 61801, USA}
\affil[33]{Dipartimento di Scienza e Alta Tecnologia, Universit\`a degli Studi dell'Insubria, via Valleggio 11, I-22100 Como, Italy}
\affil[34]{Steward Observatory, University of Arizona, 933 N. Cherry Avenue, Tucson AZ 85721, USA}
\affil[35]{Leiden Observatory, Leiden University, PO Box 9513, 2300 RA Leiden, The Netherlands}
\affil[36]{Hamburger Sternwarte, University of Hamburg, Gojenbergsweg 112, D-21029 Hamburg, Germany}
\affil[37]{Universit\"{a}t Potsdam, Karl-Liebknecht-Str. 24/25, D-14476 Potsdam, Germany}
\affil[38]{Department of Physical Sciences, Ritsumeikan University, Kusatsu, Shiga 525-8577, Japan}
\affil[39]{Academia Sinica Institute of Astronomy and Astrophysics, 11F Astronomy-Mathematics Building, AS/NTU, No.1, Section 4, Roosevelt Road, Taipei 10617, Taiwan}
\affil[40]{School of Physics and Astronomy, Tel Aviv University, Tel Aviv 69978, Israel}
\affil[41]{Kapteyn Astronomical Institute, University of Groningen, P.O. Box 800, 9700 AV Groningen, The Netherlands}
\affil[42]{Department of Physics and Astronomy, University of California Los Angeles, CA, 90095, USA}
\affil[43]{Institute for Advanced Research, Nagoya University, Furocho, Chikusa, Nagoya 464-8602, Japan}
\affil[44]{Department of Physics, Graduate School of Science, Nagoya University, Furocho, Chikusa, Nagoya 464-8602, Japan}
\affil[45]{DARK, Niels Bohr Institute, Jagtvej 155, 2200 Copenhagen N, Denmark}
\affil[46]{Institut d’Astrophysique de Paris, CNRS, Sorbonne Universit\'e, UMR7095, 98bis bd Arago, 75014 Paris, France}
\affil[47]{Department of Astronomy, University of Michigan, 1085 S. University Ave., Ann Arbor, MI 48109, USA}

\makeatletter
\renewcommand{\@maketitle}{%
{%
\thispagestyle{empty}%
\vskip-36pt%
{\raggedright\sffamily\bfseries\fontsize{20}{25}\selectfont \@title\par}%
\vskip10pt
{\raggedright\sffamily\fontsize{12}{16}\selectfont  \@author\par}
\vskip25pt%
}%
}%
\makeatother

\begin{document}

\flushbottom
\maketitle

\noindent $^*$corresponding authors\\

\noindent \textbf{Understanding the rapid formation of supermassive black holes (SMBHs) in the early universe requires insight into stellar mass growth in host galaxies. 
Here, we present NIRSpec rest-frame optical spectra and NIRCam imaging from JWST of two galaxies at $\mathbf{z>6}$, both hosting moderate-luminosity quasars.
These galaxies exhibit Balmer absorption lines, similar to low-redshift post-starburst galaxies.
Our analyses of the medium-resolution spectra and multiband photometry show  bulk of the stellar mass ($\mathbf{\log{(M_* / M_\odot) \geq 10.6}}$) formed in starburst episodes at redshift 9 and 7. 
One of the galaxies shows a clear Balmer break and lacks spatially resolved H$\alpha$ emission.
It falls well below the star formation main sequence at $\mathbf{z = 6}$, indicating quiescence.
The other is transitioning to quiescence; together, these massive galaxies are among the most distant post-starburst systems known.
The blueshifted wings of the quasar [O~{\sc III}] emission lines suggest quasar-driven outflow  possibly influencing star formation.
Direct stellar velocity dispersion measurements reveal  one galaxy follows the local black hole mass-$\sigma_*$ relation while the other is overmassive.
The existence of massive post-starburst galaxies hosting billion-solar-mass BHs in short-lived quasar phases suggests  SMBHs and host galaxies played a major role in each other's rapid early formation.}

Using the Near Infrared Spectrograph (NIRSpec) onboard the James Webb Space Telescope (JWST)~\cite{Rigby23}, we obtained rest-frame optical spectra at medium resolution of two quasars \targetone\ and \targettwo, with redshifts of 6.40 and 6.18, respectively (full coordinates provided in Methods).
The NIRSpec data of these systems with accreting SMBHs clearly show in Figure~\ref{fig:spec} (top panels) that H$\gamma$ and H$\delta$ are detected in absorption, while  H$\epsilon$ and the red part of H$\zeta$ are also present for \targetone\ due to the bluer rest-frame wavelength coverage of the spectrum.
These strong Balmer absorption lines are direct evidence for stellar emission from their host galaxies with a predominance of A- and F-type stars, as seen in post-starburst galaxies after short-lived O- and B-type stars have faded away within $\approx100$ Myr.
The NIRSpec data also reveal blueshifted broad wing components in the forbidden [O~{\sc iii}]~$\lambda5008$ emission line (Figure~\ref{fig:O3outflow}, Table~\ref{tab:Table_E1}), which are commonly taken as  evidence for ionized gas outflow on nuclear scales.
The H$\alpha$ line of \targettwo\ exhibits a double-peaked profile, which can be explained by emission from a rotating accretion disk in the quasar broad line region (see Figure~\ref{fig:J1512_Ha} and Methods for more details).

To model the underlying stellar populations in these two systems, we decomposed the spectra into quasar and host galaxy components. 
We used multiband imaging data from JWST Near-Infrared Camera (NIRCam) for this exercise.
For \targetone, we achieved contiguous coverage from 1 to 5~$\mu$m (or 1500 to 6500~\AA\ in the rest frame) with five broad-band and three medium-band filters.
For \targettwo, two-band photometry at $1.50~\mu$m and $3.56~\mu$m brackets the Balmer limit.
The stellar emission from both host galaxies was successfully detected in all NIRCam images by performing a two-component model fit which includes an unresolved quasar and an extended host galaxy (Figure~\ref{fig:NIRCam_decomposition}  and  Table~\ref{tab:Table_E2}).
Figure~\ref{fig:img_slit} shows that the host galaxies span the $0\farcs2$-wide NIRSpec slit in the dispersion direction.
Nonetheless, they are intrinsically compact, with effective radii measured along the semi-major axis no greater than 1~kpc (Table~\ref{tab:Table_E3}).
In this decomposed multiband photometry, the clear photometric break at rest-frame $\approx4000$~\AA\ and the lack of an excess in F480M, where H$\alpha$ falls at $z\sim6.4$, provide further evidence that \targetone\ has not been forming stars recently (Figure~\ref{fig:spec}, bottom left panel).
Note that these two features are sensitive to different timescales~\cite{Kennicutt12}, with the break tracing the past 100~Myr of star formation and the H$\alpha$ emission tracing the past 10~Myr.
There is little spatially extended H$\alpha$ emission in the 2D spectrum of \targetone\ (C.~Phillips et al., manuscript in preparation), consistent with the flat F444W $-$ F480M color of the host.
We first model the quasar continuum shape and subtract the  continuum plus  emission lines to isolate the host galaxy spectra (top panel of Figure~\ref{fig:spec}), as detailed in Methods.
We then scale the extracted host spectra to match the decomposed photometry of the host galaxies at 3.56 $\mu$m, accounting for 
  the flux loss from the $0".2$-wide slit.

We then fit the decomposed host galaxy photometry and spectra with stellar population models using \textsf{Bagpipes}, a public spectral fitting code~\cite{Carnall18, Carnall19}. 
We use the Kroupa initial mass function (IMF)~\cite{Kroupa01} and two models for star formation history (SFH): a delayed-$\tau$ model where the star formation rate SFR($t$) $\propto t \exp{(-t/\tau)}$ with $\tau$ denoting the timescale of the exponential decline of star formation, and a non-parametric SFH model with a continuity prior presented in ref~\cite{Leja19}.
We use seven time bins for the non-parametric model with bin edges at 0, 10, 100, 200, 300, 400, 600 and 800 Myr before the redshift of observation.
The best-fit SED model and the recovered SFHs for each target are shown in the bottom panels of Figure~\ref{fig:spec}. 
The posteriors of the model parameters are given in Table~\ref{tab:Table_E4} and Figure~\ref{fig:corner_J2236}.

In the delayed-$\tau$ SFH model, we find that the stellar masses of $\log{~M_* / M_\odot} = 10.80_{-0.02}^{+0.03}$ ($\pm0.08$) for \targetone\ and $\log{~M_* / M_\odot} = 10.64_{-0.01}^{+0.04}$ ($\pm0.02$)  for \targettwo~ were built up in bursts of star formation as shown by the rapid rise and fall of their inferred star formation history (Fig.~\ref{fig:spec}, bottom right).
Here we report the errors from model fitting and, in parentheses, those due to the imaging decomposition in the F356W filter (18\% for \targetone, and 6\% for \targettwo).
Additional uncertainties due to the assumptions of the IMF, the  radial profile  (see Methods), and biases in the assumed SFH shapes are not taken into account.
The non-parametric SFH model returns stellar masses consistent with the delayed-$\tau$ SFH model for both targets.

The recovered SFHs suggest that the two galaxies experienced starburst episodes at earlier times.
The delayed-$\tau$ SFH model indicates that \targetone\ had a peak SFR of $1910^{+870}_{-620}\ M_\odot$ yr$^{-1}$ with a mass-weighted stellar population age of $270^{+20}_{-30}$ Myr.
This age, corresponding to a formation redshift of $z_\mathrm{form}=8.6^{+0.3}_{-0.4}$,
represents the time when the half of the observed stellar mass was formed. 
Similarly, the inferred peak SFR of \targettwo\ is $1400^{+520}_{-430}\ M_\odot$ yr$^{-1}$ with the mass-weighted age of $140^{+30}_{-10}$ Myr ($z_\mathrm{form}=7.1_{-0.1}^{+0.2}$).
The non-parametric SFH model yields less extreme yet high SFR peaks with $300^{+140}_{-70}\ M_\odot$ yr$^{-1}$ for \targetone\ and $480^{+100}_{-120}\ M_\odot$ yr$^{-1}$ for \targettwo.
The inferred age is 40--130 Myr longer than the delayed-$\tau$ model.
The difference between the two SFH models is attributed to the distinct SFH shape assumed by each SFH model and also to the fact that the present data are not sensitive to stellar populations older than those observable in the rest-frame optical stellar continuum.

In contrast to its previous vigorous starburst activity, the SFR of \targetone\  dropped sharply, with an $e$-folding timescale of $\tau=18^{+9}_{-5}$ Myr.
The non-parametric SFH model also indicates a $\gtrsim2$ dex decline of SFR over the past 100 Myr.
As a result, both SFH models confirm that the $2\sigma$ upper limits of specific SFR, defined as the ratio of SFR to stellar mass,  is  $\lesssim 0.1$ Gyr$^{-1}$ and $\lesssim 0.01$ Gyr$^{-1}$  when averaged over the past 100~Myr and 10~Myr, respectively.
This galaxy was quenched approximately 100~Myr before the observation, corresponding to a quenching redshift $z_\mathrm{quench} = 7.6_{-0.2}^{+0.3}$ (delayed-$\tau$ SFH) or $z_\mathrm{quench} = 7.1$ (non-parametric SFH), where the quenching time is the epoch when the SFR times the Universe's age drops below 20\%\ of the stellar mass~\cite{Pacifici16}.
This rapid quenching is consistent with what has been found for other high-redshift massive quiescent galaxies~\cite{Carnall23, DeGraaff24, Weibel24}.

\targettwo, which experienced its peak star formation more recently, is currently on the star-formation main sequence, but it is transitioning to quiescence (Figure~\ref{fig:quiescence}, left) with an $e$-folding time similar to \targetone. 
Both SFH models indicate that the specific SFR decreases from $\lesssim1$~Gyr$^{-1}$ (averaged over the past 100~Myr, $2\sigma$) to $\lesssim0.2$~Gyr$^{-1}$ (10~Myr).
The latter places \targettwo\  around the threshold for quiescence.
We conclude that both  $z>6$ quasar host galaxies are among the earliest massive post-starburst galaxies known, rapidly quenching within the first billion years of the universe (Figure~\ref{fig:quiescence}, right).

We can now chart the mass assembly histories of these two quasar host galaxies and their central SMBHs as a function of cosmic time. 
As illustrated in the left panel of Figure~\ref{fig:J2236t}, the two galaxies in this study had already grown to stellar masses of $10^{10.5}~M_\odot$ by redshifts 9 and 7, possibly suggesting that they are the evolved descendants of the currently known highest-redshift galaxies hosting AGN~\cite{Maiolino24, Goulding23}.
The rapid stellar mass growth of \targetone\ and \targettwo\ stands in contrast to the smooth mass growth predicted by theoretical models\citep{Zhang24, Scoggins24}.

Another challenge is to explain how the SMBHs that power these quasars formed in the early universe.  
These quasars have estimated black hole masses of 
$M_\mathrm{BH} = 1.1 \times 10^9~M_\odot$ and $1.3 \times 10^9~M_\odot$, based on broad Balmer emission lines (see Methods). 
If these started as seed BHs growing constantly at the Eddington limit, the seed mass is $\lesssim 10^3~M_\odot$, depending on the seeding epoch (middle panel).
This scenario, however, is not strongly preferred, because we estimate sub-Eddington accretion rates at the observed redshifts.
An alternative and probably more realistic scenario involves intermittent super-Eddington episodes, as demonstrated in semi-analytical calculations by ref~\cite{LiW23} (Figure~\ref{fig:J2236t}, grey lines in the middle panel).
This evolutionary scenario, in which the exponential BH accretion mode turns on and off for short periods of time, is generally consistent with studies which suggest that the duty cycle of $z>6$ quasars — the fraction of time when BH accretion is active — is significantly smaller than unity \citep{Davies19, Eilers24}.

The right panel of Figure~\ref{fig:J2236t} illustrates the evolution of the two galaxies in the $M_*$ -- $M_\mathrm{BH}$ plane, in which we assume constant Eddington limit accretion throughout for simplicity.
The suppressed stellar mass growth causes the two SMBHs, now observed  above the local bulge mass - SMBH mass relation (grey line), 
to climb up the co-evolution plane after the starburst phase.
This scenario holds unless the SMBHs reached $10^9 M_\odot$ by the starburst epoch, which is unlikely given the young age of the universe and the evidence that rapid SMBH assembly typically follows a host starburst in the local universe ~\cite{Davies07, Wild10}.
These galaxies may also have experienced SFH similar to those of  known $4<z<11$ JWST AGN~\cite{Maiolino24b, Goulding23, Maiolino24}, although  the pre-starburst SFHs of our sample remain poorly constrained.
We also point out that further observations of the local environments are required to address whether these galaxies have completely halted star formation, or whether they will be rejuvenated once the gas expelled by the quasar feedback  falls back or is refilled from their host halos~\cite{Lupi24}.

The clear detection of stellar absorption lines allows us to measure the stellar velocity dispersion $\sigma_*$, a tracer of the depth of the central mass potential.
We measure $\sigma_*$ for the two galaxies as part of the Bagpipes SED fitting analysis, in which the velocity profiles of the stellar population models are convolved with Gaussian kernels to fit the observed Balmer absorption lines.
From the delayed-$\tau$ SFH model, we recover stellar velocity dispersions of the Balmer lines of $\sigma_*=290_{-60}^{+50}$~km s$^{-1}$ for \targetone\ and $\sigma_*=160_{-40}^{+30}$~km s$^{-1}$ for \targettwo.
These are the first direct measurements of stellar velocity dispersion for quasar hosts at $z>6$.
They are also among the most distant successful measurements of stellar absorption lines for any high-redshift galaxy to date~\cite{Tanaka19, Looser24, Weibel24}.
We also use \textsf{pPXF}, a widely-used tool for measuring  $\sigma_*$, to validate the \textsf{Bagpipes} results  (see Methods and Figure~\ref{fig:ppxf}).
For \targetone, the inferred $\sigma_*$ from \textsf{pPXF} agrees with that from  \textsf{Bagpipes}.
For \targettwo, however,  the inferred $\sigma_*$  is consistent with an unresolved line within the $1\sigma$ uncertainty.
We therefore adopt a conservative $2\sigma$ upper limit of $\sigma_* < 190$ km s$^{-1}$, as reported in Table~\ref{tab:Table_E4}.
Note that the spectrum of \targettwo\ covers only the central region of the host galaxy, which is smaller than the effective radius along the semi-major axis~(Figure~\ref{fig:img_slit}), whereas for \targetone\ the slit is aligned with the orientation of the host galaxy.
This may affect the interpretation of the reported $\sigma_*$ constraints.

We show the distribution of the black hole mass and  $\sigma_*$ in Figure~\ref{fig:Msigma}, along with local samples and the result for GS-9209, a $z=4.7$ quiescent galaxy hosting an accreting black hole~\cite{Carnall23}.
\targetone\ falls on the local relation of ref~\cite{KorHo13}, while \targettwo\ lies  slightly above it, suggesting that the tight $M_\mathrm{BH}$--$\sigma_*$ correlation observed in the local universe is  being established at $z\sim6$, with a larger  scatter at earlier times.
The right panel of Figure~\ref{fig:Msigma} shows that the ionized gas velocity of [O~{\sc iii}] does not necessarily trace  $\sigma_*$, highlighting the need for direct  $\sigma_*$ measurements of high-redshift AGN to characterize the redshift evolution of the underlying $M_\mathrm{BH}$--$\sigma_*$ relation.

The existence of massive quenched galaxies at high redshifts~\cite{Carnall23, Looser24, DeGraaff24}, as seen by JWST, challenges our understanding of galaxy evolution.
While a tight connection between starburst and AGN activity has been proposed for decades~\cite{Davies07, Wild10}, it has been unclear whether quasar activity is the primary cause of star formation quenching. 
Our discovery of post-starburst galaxies hosting quasars within the first 900 Myr of the universe  adds a layer of intrigue. 
These observations reveal that, in the high-redshift gas-rich universe, galaxies can undergo a rapid transition from a starburst phase to a quiescent state, while their central SMBHs continue to accrete gas and appear as quasars.
Similar cases have recently been reported~\cite{Ito22, Carnall23, Belli24, D'Eugenio24} at $2<z<5$.
\targetone\ and \targettwo\ may represent a brief (a few $100$~Myr) phase of quasar-driven galaxy quenching or it could also be the cumulative effect of multiple quasar episodes associated with the earliest-formed SMBHs that quenched star formation.
The latter interpretation aligns with the short quasar lifetime suggested in quasar clustering analyses ($\approx1$--$10$~Myr)~\cite{Arita23, Eilers24} and  theoretical models of early SMBH growth~\cite{LiW23} (see Figure~\ref{fig:J2236t}, left panel).
We point out that the high peak SFRs inferred for \targetone\ and \targettwo\ are comparable to the host SFRs of UV luminous quasars in the same redshift range (SFR~$\approx 100$--$1000\ M_\odot\ \mathrm{yr^{-1}}$)~\cite{Decarli18, Walter22}, implying that these two galaxies are in a later quenching phase than the starbursting quasar hosts previously known.
Our results shed light on the mechanisms driving quasar activity and its role in galaxy quenching at early cosmic epochs, highlighting the need for a better understanding of the mutual growth history of SMBHs and their host galaxies.


\newpage
\section*{Figures}
\begin{figure}[h]
\centering
 \includegraphics[width=0.8\linewidth]{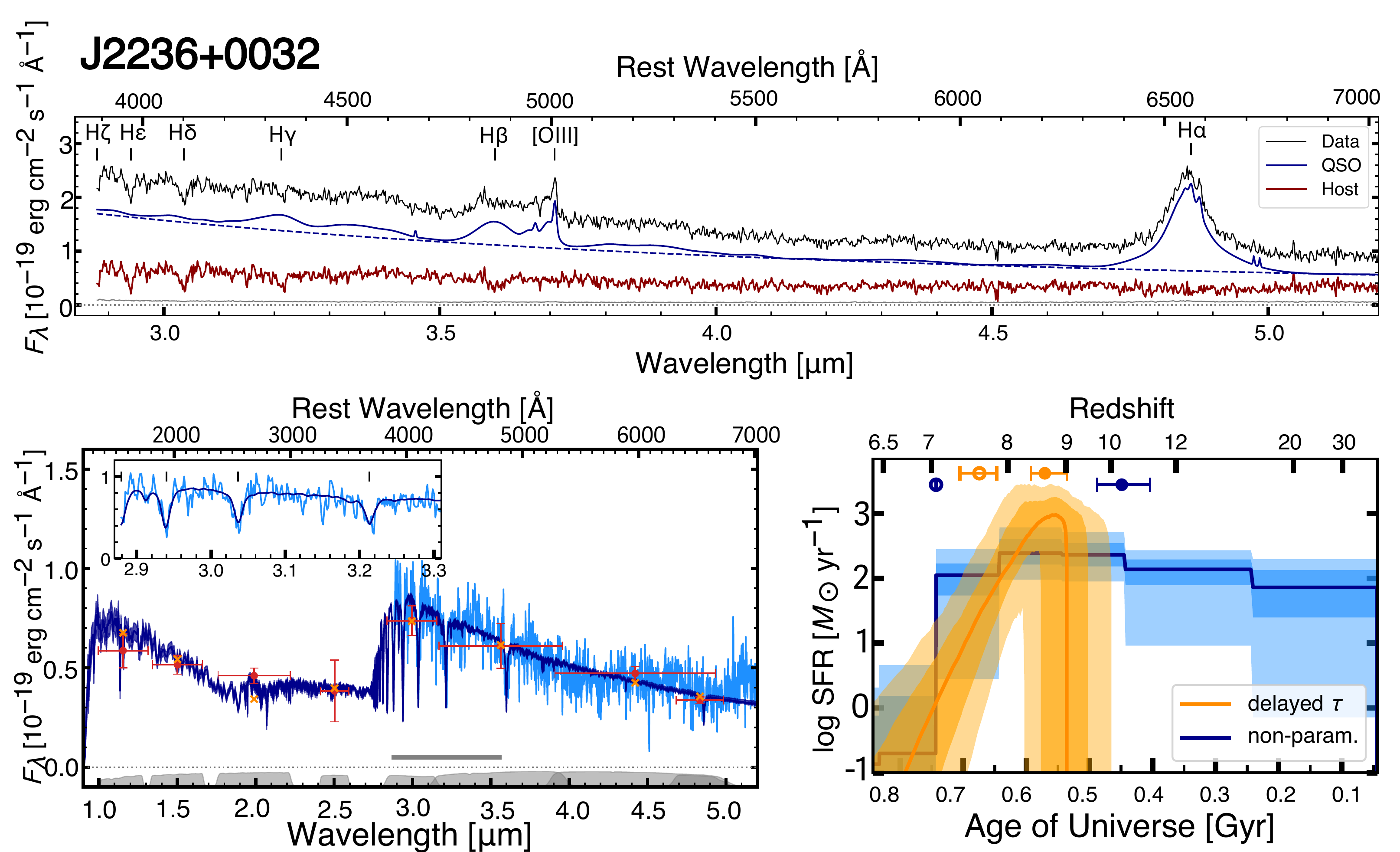}
 \includegraphics[width=0.8\linewidth]{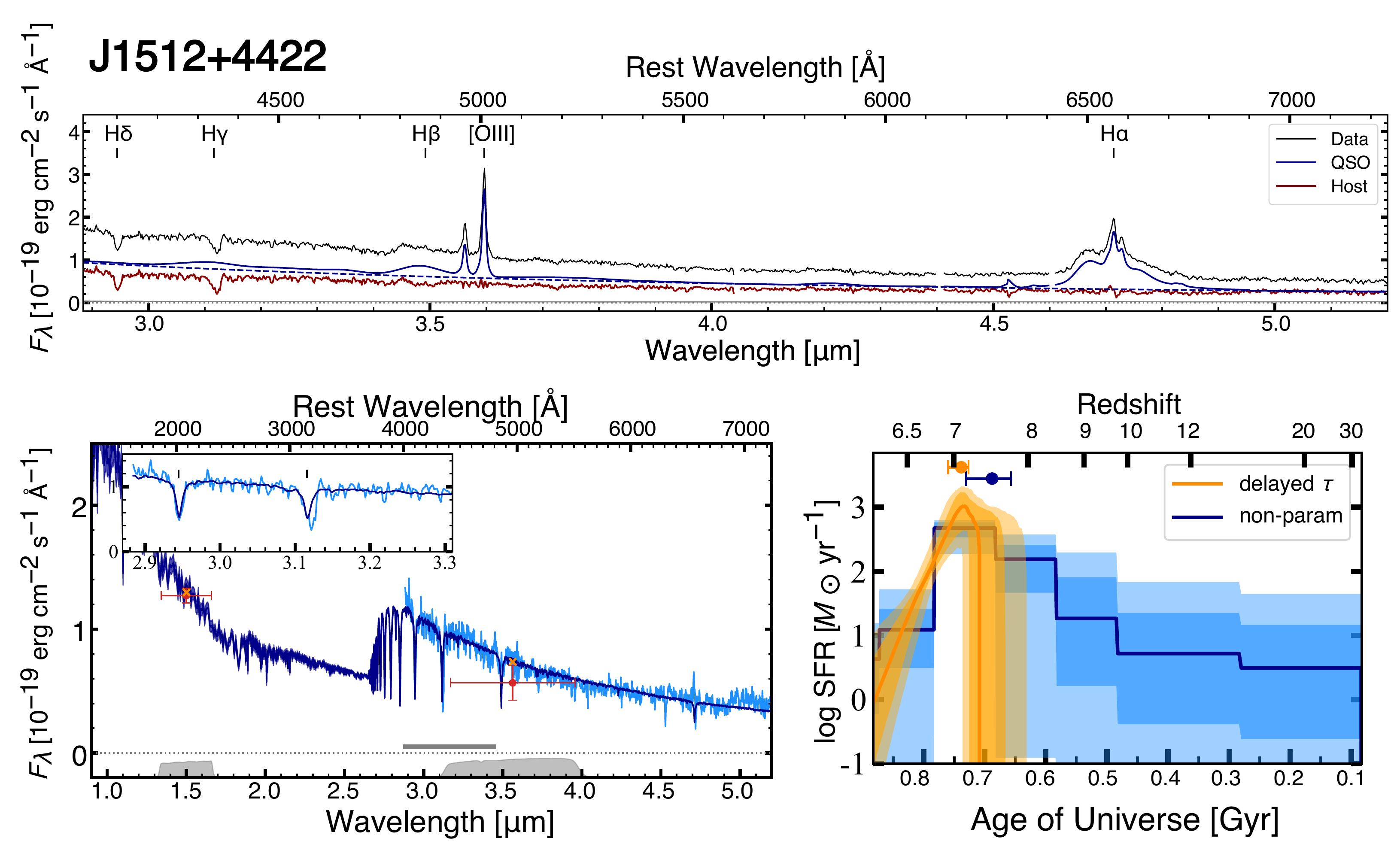}
 \vspace{-0.2cm}
\caption{\textbf{JWST NIRSpec G395M spectra of \targetone\ and \targettwo.}
Balmer absorption lines, indicative of a post-starburst phase, are clearly detected.
(\textit{Top:}) The decomposed quasar and host galaxy components are shown in blue (dashed: power-law continuum; solid: continuum plus emission line model) and red, respectively. 
The grey solid line represents the error spectrum.
Pixels affected by cosmic rays are masked.
(\textit{Bottom left:}) Best-fit galaxy SED models from \textsf{Bagpipes} according to the delayed-$\tau$ SFH model (dark blue). 
The extracted host galaxy spectrum, scaled to match the decomposed F356W host photometry, is shown in cyan.
The grey bar at the bottom shows the wavelength range of the spectrum used in the SED fit.
NIRCam photometry is shown with the red symbols with the filter transmission curves displayed at the bottom. 
Error bars in the x and y directions represent the effective bandwidth and the photometric uncertainties, respectively.
Inset panels highlight the Balmer absorption lines.
(\textit{Bottom right:}) Recovered SFHs  (delayed-$\tau$: orange, non-parametric: blue).
Solid lines show median posteriors; darker (lighter) shaded regions indicate 16th--84th (2th--98th) percentile intervals.
Filled and open symbols indicate the median posteriors of  formation redshift ($z_\mathrm{form}$) and quenching redshift ($z_\mathrm{quench}$), respectively, for each SFH model. 
Error bars indicate the 16th–84th percentile ranges.
} 
\label{fig:spec}
\end{figure}

\begin{figure}[p]
\centering
 \includegraphics[width=\linewidth]{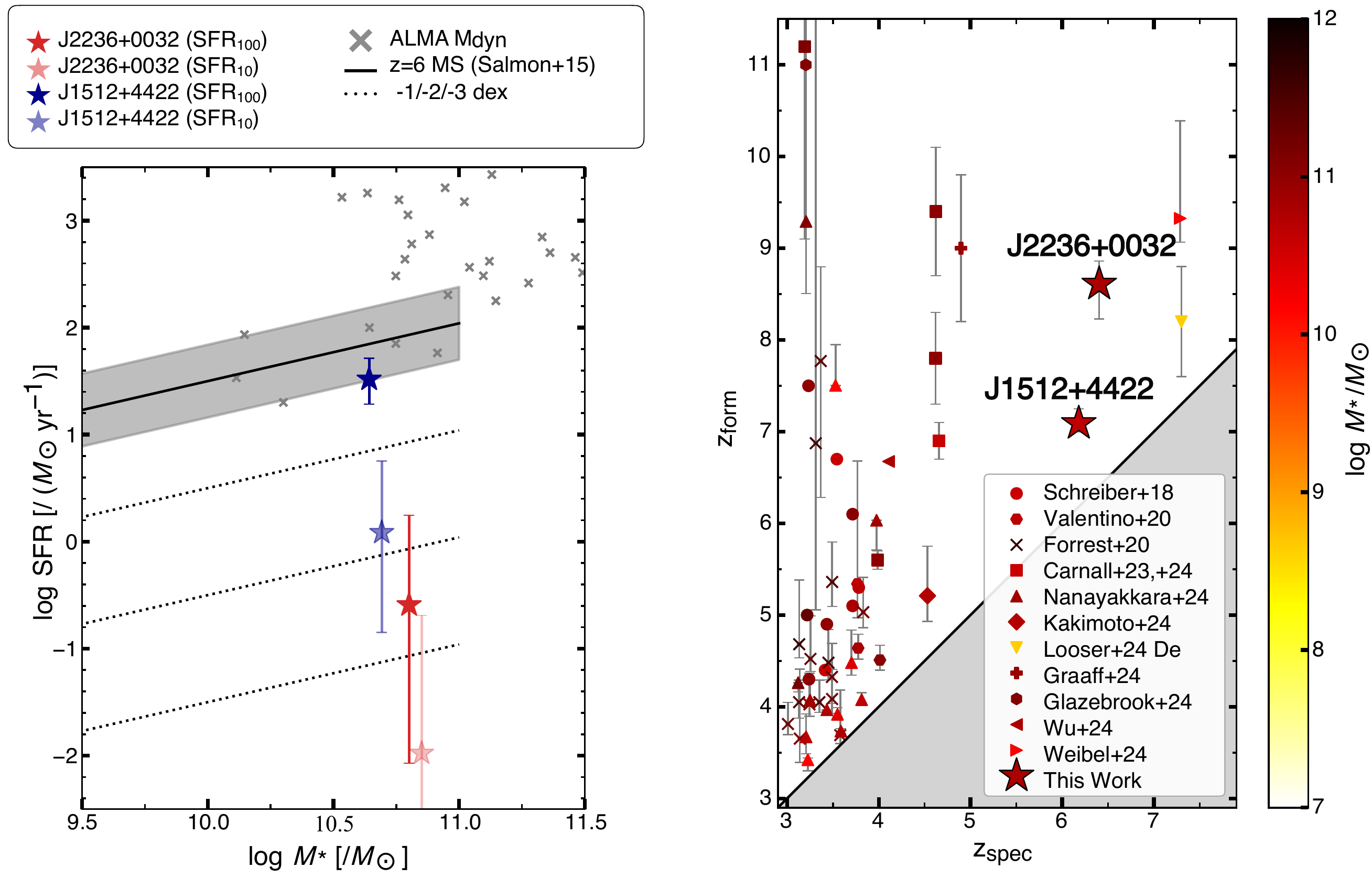}
\caption{ \textbf{Quiescence of the two quasar host galaxies.}
(\textit{Left:}) The stellar mass -- star formation rate distribution of quasar host galaxies at $z\gtrsim6$, compared with the star formation main sequence~\cite{Salmon15} at $z\sim 6$ and its scatter (black dashed line and grey shading).
The red and blue symbols represent the median posteriors of the inferred SFR (delayed-$\tau$ SFH model) for \targetone\ and \targettwo, respectively.
For each target, the SFR averaged over the last 100~Myr and 10~Myr is shown as darker and lighter colors, respectively.
Error bars indicate the 16th–84th percentile ranges.
ALMA dynamical mass measurements and obscured SFR for $z\gtrsim6$ quasars compiled in ref~\cite{Izumi21b} are shown as grey crosses.
The dotted lines show the star formation main sequence offset by $-1$, $-2$, and $-3$ dex, respectively.
(\textit{Right:}) Spectroscopic redshift $z_\mathrm{spec}$ and formation redshift $z_\mathrm{form}$ of known quiescent galaxies at $z>3$, color-coded by stellar mass (more massive galaxies are shown in redder colors).
The targets of this work (the delayed-$\tau$ SFH model) are the most distant such objects  known with stellar mass $\log{M_* / M_\odot} > 10$.
Comparison data are from refs\cite{Schreiber18, Valentino20, Forrest20, Carnall23, Carnall24, Nanayakkara24, Kakimoto24, Looser24, DeGraaff24, Glazebrook24, WuPF25, Weibel24}.
Symbols and error bars indicate best-fit values and their 1$\sigma$ ranges.
We note that several other galaxies with Balmer breaks have been reported at high redshift\cite{WangB24, Kokorev24, Labbe24}.
} \label{fig:quiescence}
\end{figure}


\begin{figure*}[htbp!]
\centering
 \includegraphics[width=\linewidth]{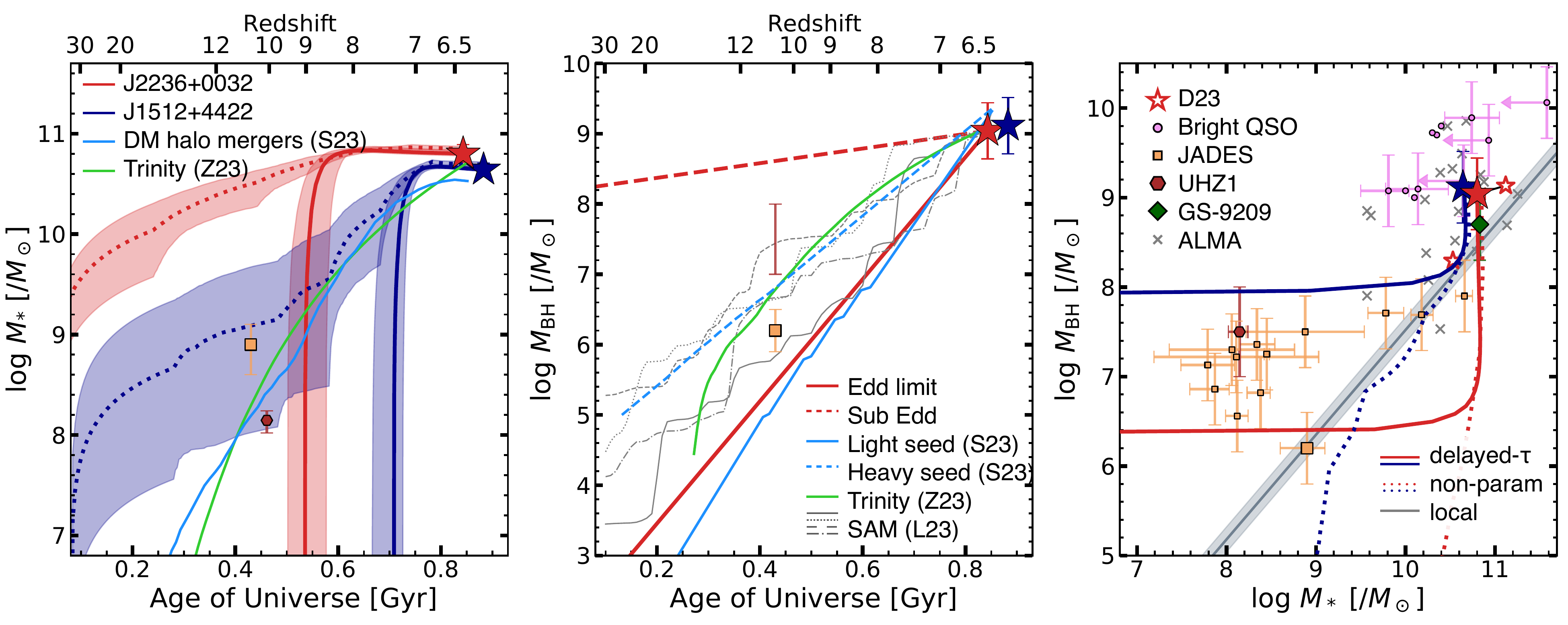}
 \caption{\textbf{Growth pathways of the two quasar host galaxies and their SMBHs.} 
 (\textit{Left:}) Stellar mass assembly as a function of cosmic time based on the delayed-$\tau$ SFH models (solid lines) and non-parametric SFH models (dotted lines). 
The 16th--84th percentile range is indicated by the shaded region for each model.
For comparison, the average growth curve from the \textsf{Trinity} simulation~\cite{Zhang24} is shown in green, and that based on the dark matter halo merger history from ref~\cite{Scoggins24} is shown in cyan. 
Both model curves, fine-tuned to reproduce the observed stellar mass of \targetone, are smooth and in contrast to its inferred mass growth, which reaches $>10^{10.5}~M_\odot$ by $z\sim8$.
Two of the known $z>10$ galaxies hosting AGN are also presented (GN-z11~\cite{Maiolino24} in orange; UHZ1~\cite{Goulding23} in brown).
(\textit{Middle:}) BH mass assembly as a function of time.  The red line show cases in which \targetone\ continuously accretes at the Eddington limit (solid) or at the observed Eddington ratio ($10\%$; dashed) both with $10\%$ radiative efficiency.
The blue lines show the models of ref~\cite{Scoggins24}, where the seed BH of \targetone\ is a light seed of $50 M_\odot$ (solid line), or a heavy seed of  $10^5 M_\odot$ (dashed line). 
The four grey lines show the semi-analytical model (SAM) from ref~\cite{LiW23},  where seed BHs grow by changing Eddington ratios (including super-Eddington accretion rates) for short periods of time.
The four BHs presented here are a representative sample of BHs that reach $\sim 10^9 M_\odot$ by $z=6$. 
These model growth curves are scaled to match the BH mass of \targetone.
(\textit{Right:}) Evolution in the $M_\mathrm{BH}$ -- $M_*$ plane (solid line: delayed-$\tau$ SFH, dotted line: non-parametric SFH model). 
Luminous quasars whose host stellar emission is detected with JWST~\cite{Stone23, Stone24, Yue24}  (pink) and faint  AGN from the JWST JADES program~\cite{Maiolino24b} (orange, with smaller symbol sizes than GN-z11)  are shown for comparison. 
Quasar systems with ALMA dynamical masses and Mg{\sc ii}-based $M_\mathrm{BH}$  compiled by ref~\cite{Izumi21b} are shown as grey crosses.
The $z=4.7$ quiescent galaxy GS-9209~\cite{Carnall23} is shown in green. 
The stellar and BH masses of the two $z>6$ quasar host galaxies in the earlier study of ref~\cite{Ding23}, which includes \targetone, are shown as open red stars.
The local bulge mass -- BH mass relation from ref~\cite{KorHo13} is shown as the grey line.
Error bars represent the $1\sigma$ ranges in all panels.
} \label{fig:J2236t}
\end{figure*}

\begin{figure*}[btp!]
\centering
 \includegraphics[width=\linewidth]{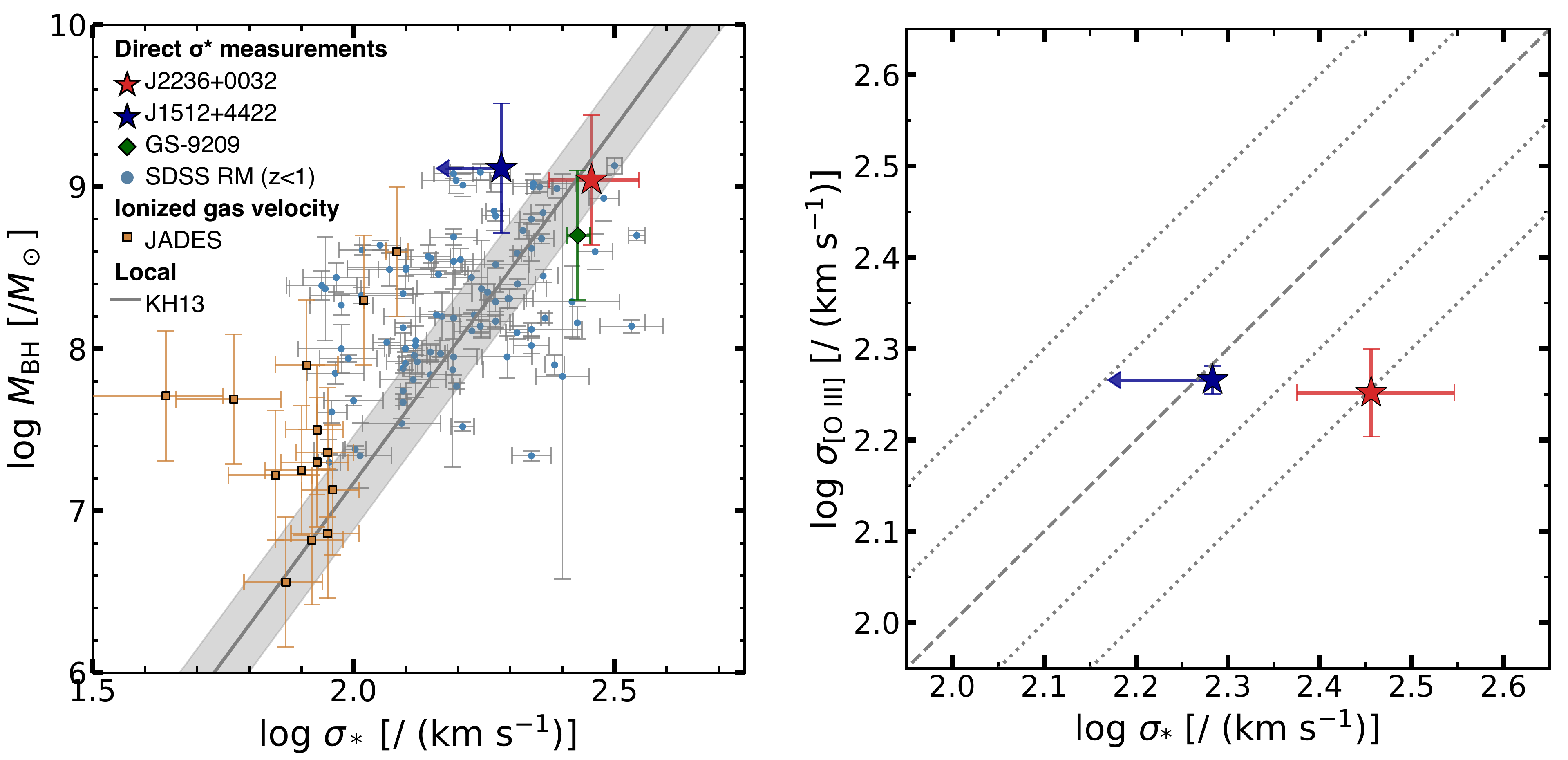}
 \caption{ \textbf{Stellar velocity dispersion ($\sigma_*$) measurements.}
 (\textit{Left:})  
Distribution of  $M_\mathrm{BH}$ versus  $\sigma_*$.
Blue and red symbols represent \targetone\ and \targettwo, respectively.
The green symbol shows GS-9209~\cite{Carnall18}, and orange symbols show the JADES sample~\cite{Maiolino24b}.
Cyan symbols show the $z<1$ quasars from the SDSS reverberation mapping project~\citep{Matsuoka15}.
Error bars represent the $1\sigma$ ranges, except for $\sigma_*$ of \targettwo, which is shown as a $2\sigma$ upper limit (see text for details).
Note that $\sigma_*$ measurements are based on stellar absorption lines for our targets and GS-9209, while the JADES sample use the ionized gas velocity dispersion measured from the [O~{\sc iii}] emission lines ($\sigma_\mathrm{[O~III]}$).
The grey line and shading show the local relation and its scatter~\cite{KorHo13}. 
(\textit{Right:}) Comparison of $\sigma_*$ estimated from the Balmer absorption lines with $\sigma_\mathrm{[O~III]}$.
The diagonal dashed line is the one-to-one relation, while the dotted lines indicate the case where the [O~{\sc iii}] velocity dispersion  is offset from the spectroscopic $\sigma_*$  from 
$-0.2$ to $+0.2$ dex, in steps of 0.1 dex.
For \targetone, $\sigma_*$ is larger than $\sigma_\mathrm{[O~III]}$ by $\approx +0.2$ dex.
} \label{fig:Msigma}
 \end{figure*}

\clearpage

\noindent \textbf{\Large Methods}

\medskip
\noindent {\bf Cosmological model}  \label{sec:cosmology}

\noindent A standard  cosmology with $H_0= 70$~km s$^{-1}$ Mpc$^{-1}$, $\Omega{_m} = 0.30$, and $\Omega{_\Lambda} = 0.70$ is adopted, which gives a scale of $5.51$ and $5.62$ proper kpc per arcsecond at $z=6.4$ and $z=6.18$, respectively. All magnitudes are presented in the AB system. 
Milky Way dust extinction is negligible at the near-infrared wavelengths of interest and is not corrected for.

\medskip
\noindent {\bf Spectroscopic data}

The targets presented in this paper, HSC~J223644.58$+$003256.9~\cite{Matsuoka16} at $z=6.4$ (hereafter \targetone) and HSC~J151248.71$+$442217.5~\cite{Matsuoka19b} at $z=6.18$ (hereafter \targettwo) were originally discovered by the Hyper Suprime-Cam Subaru Strategic Program (HSC-SSP~\cite{HSC-SSP2018}). 
Their HSC $y$-band magnitudes and absolute ultraviolet magnitudes at rest-frame 1450 \AA\ are ($y$, $M_{1450}$) = (23.19, -23.75) and (24.16, -22.07), respectively.
These two quasars were observed in a JWST Cycle 1 program using NIRSpec and NIRCam (GO 1967; PI: M.Onoue), which aims to characterize the properties of the central SMBHs and host galaxies of the moderate-luminosity HSC-SSP quasars.
The [O~{\sc iii}] redshifts of the two targets, based on their narrow components, are consistent with the rest-frame ultraviolet measurements ($|z_\mathrm{UV} - z_\mathrm{[OIII]}| \leq 0.01$; Table~\ref{tab:Table_E1}), as detailed below.

The JWST NIRSpec observations were executed on 2022 October 28 for \targetone\ and 2023 February 14 for \targettwo. The targets were aligned onto the $0\farcs2$-wide S200A2 slit using the wide aperture target acquisition (WATA) method. Data were taken with the medium-resolution G395M grating, which covers 2.87--5.27~$\mu$m. 
For our emission and absorption line measurements, we use the spectral resolution of $R = 750$, which is provided in the JWST documentation.
Note that both host galaxies are extended along the dispersion direction of the S200A2 slit, which is $0\farcs2$ wide, as shown in Figure~\ref{fig:img_slit}.
The exposure for \targetone\ was divided into 45 groups at each of three primary dither positions without subpixel dithers, amounting to 1,970 seconds. 
The NRSIRS2RAPID mode was used for detector readout.
The $y$-band magnitude of \targettwo\ is 1 magnitude fainter than \targetone, so we integrated longer, using the  NRSIRS2 detector readout mode.
The target was observed with 18 groups each at three primary dither positions, 
which was repeated twice to  achieve a total exposure time of 7,878 seconds.

Here we briefly describe the NIRSpec data reduction in this paper; the details are given in ref~\cite{Ding23}.
The raw (\textsf{uncal}) spectroscopic data were downloaded from the MAST archive and processed with JWST pipeline version 1.17.1 with parameter reference files \textsc{jwst\_1100.pmap} (\targetone) and \textsc{jwst\_1069.pmap} (\targettwo), as registered in the \jw\ Calibration Reference Data System (\url{https://jwst-crds.stsci.edu}).
The 1/f noise in the 2D spectra is subtracted after the Stage~1 pipeline reduction by using a public code implemented in \textsf{msaexp}~\cite{msaexp}. 
For each target, the point-source pathloss correction was applied during the Stage~2 pipeline reduction to make sure that the quasar light from the targets is fully flux-calibrated.
The processed two-dimensional \textsf{cal} spectra at each dither position were stacked using the Stage~3 pipeline with inverse-variance weighting.
One-dimensional spectra were extracted with a 6-pixel-wide  (0\farcs{}6 wide) box-car aperture.

\medskip
\noindent {\bf Photometric data}

JWST broadband NIRCam imaging data were also available for the two targets.
Images of two broadband filters (F150W and F356W) were obtained for each target in GO \#1967.
These two filters were chosen to straddle the redshifted Balmer break.
\targetone\ was also observed in an additional six photometric filters (F115W, F200W, F250M, F300M, F444W, and F480M) in Cycle 2 (GO \#3859; PI: M.Onoue).
These eight-band photometric data cover the rest-frame wavelength range from 1550~\AA\ to 6530~\AA.
The H$\alpha$ line of \targetone\ falls in the medium-resolution F480M filter, which makes it possible to address the H$\alpha$ emission strength of the host galaxy.
The raw imaging data were processed with the same procedure as in ref~\cite{Ding23} with the  pipeline version 1.8.5.
For both the Short Wavelength (F115W, F150W, F200W) and Long Wavelength (F250M, F300M, F356W, F444W, F480M) data, single visit images from the Stage~2 pipeline were stacked with a pixel scale a factor of two smaller than that of the detector during the  \textsc{Resample} step of the Stage~3 pipeline processing.
The final pixel scales are $0.0153''$ for SW and $0.0315''$ for LW images, respectively.
The astrometry is calibrated against GAIA DR3 stars within the field of view of the target images.

\medskip
\noindent {\bf 2D image decomposition for NIRCam data}

We use \textsf{galight}~\cite{Ding20}, which utilizes the image modelling capabilities of \textsf{lenstronomy}~\cite{lenstronomy} to separate the bright quasar light from the stellar light of the underlying host galaxy. 
A scaled point spread function (PSF) is used to represent the unresolved quasar emission.
The radial profile of the stellar emission from the host galaxy is modeled by a 2D S\'ersic profile convolved with the PSF model.
This software has been tested with many astronomical images from the Hubble Space Telescope~\cite{Ding20} and the ground-based Subaru/HSC~\cite{LiJ21}.
\citetalias{Ding23} shows that the stable and sharp JWST PSF makes it possible for \textsf{galight} to detect quasar host galaxies~\cite{Ding23, Tanaka24} at $z=2$--$6$.
The shape of the PSF, which is crucial in the decomposition analysis, is based on bright isolated stars detected in the same image as the quasars.
We apply each of these PSF models in turn in fitting  the target images, and select the top five PSF models based on $\chi^2$.
The imaging decomposition results reported in this work are based on the best-fit model with their uncertainties derived from the dispersion in the host properties from these five PSF models.
During the fitting process, we use a fixed S\'ersic index of $n=3$ for the two target galaxies.
The choice of this S\'ersic index is motivated by a morphology study of high-redshift quiescent galaxies from ref~\cite{Ito24}.
We also performed the imaging decomposition analysis with the S\'ersic index fixed to $n=1, 2$, and $4$; however, we find no significant difference in the goodness-of-fit between the four cases because of the degeneracy between the unresolved quasar emission and the host emission, especially given how compact the host galaxies of our two targets are.
The imaging decomposition analysis is performed separately for the images in each of the eight (\targetone) and two bands (\targettwo).

The results of the 2D image decomposition are reported in Table~\ref{tab:Table_E2}. 
We successfully recovered the host stellar emission from both targets across all filters. 
The original NIRCam images and the host images after PSF subtraction are shown in Figure~\ref{fig:NIRCam_decomposition}.
\targetone\ and \targettwo\ show the brightest stellar emission in F356W among the twelve targets observed in our Cycle~1 program. 
Table~\ref{tab:Table_E3} shows that both host galaxies have compact morphologies in F356W images, with effective radii of 0.10--0.17 arcseconds, corresponding to $R_\mathrm{eff} = 0.55$--$0.96$ proper kpc.  
The $R_\mathrm{eff}$ values measured in other filters are also  below 1~kpc, supporting the robustness of the fitting results.  
In addition to their compactness, these galaxies exhibit moderately elongated shapes, with minor-to-major axis ratios of $q = 0.3$--$0.5$.
The full results for our Cycle~1 and 2 data will be presented in ref~\cite{Ding25}.

\medskip
\noindent {\bf Spectroscopic decomposition for the NIRSpec data}

We decompose the NIRSpec rest-frame optical spectra into quasar and host components.
We first perform continuum+emission line spectral fitting for the obtained spectra in order to model strong emission lines, a power-law quasar continuum, and the iron pseudo-continuum, which should be subtracted for the host galaxy analysis.
Note that stellar emission is not taken into account in this step.
For this purpose, we run a public spectral fitting tool \textsf{QSOFitMORE} \citep{Fu21} (version 1.2.0\footnote{\url{https://doi.org/10.5281/zenodo.4893646}}) with custom modification to fit the continuum and emission lines simultaneously.
We manually mask pixels which are significantly affected by the high-order Balmer absorption lines,
allowing us to fit broad emission components for H$\gamma$ (\targetone\ and \targettwo) and H$\delta$ (\targettwo).
Each of these two Balmer emission line components is fitted with a single Gaussian profile.
As is also described by \citetalias{Ding23}, \targetone\ needs one Gaussian profile for H$\beta$, while one narrow and one broad components for each of the [O~III] doublet line.

To subtract the quasar continuum emission from the line-free spectra, we adopt a single power-law function, $f_\lambda \propto \lambda^{\alpha_\lambda}$, where $\alpha_\lambda$ represents the continuum slope.  
For \targetone, we fit a single power-law function to the decomposed quasar NIRCam photometry in the three filters that encompass the NIRSpec data (F300M, F356W, and F444W).  
The host galaxy spectrum is extracted by subtracting the emission-line (including iron) plus power-law quasar model from the original spectrum.  
To account for extended emission falling outside the S200A2 slit, we rescale the decomposed host spectrum upward by 7\% to match the host galaxy photometry in F356W.  
After performing the initial host SED fitting analysis, we iterate the emission-line fitting and host extraction, using the best-fit host SED model, as detailed later in this Methods section. 
In the second iteration, we model the line and continuum emission  using the spectrum from which the galaxy model has been subtracted.  
We find that this iterative procedure improves the modeling of the broad Balmer lines, which are affected by stellar absorption.  
The resulting quasar continuum slope is $\alpha_\lambda = -1.89 \pm 0.26$.  
The host SED properties reported in this paper are based on the spectrum with the quasar continuum subtracted using this updated model.

For \targettwo, we initially fix $\alpha_\lambda$ to $-1.7$, a typical value for low-redshift quasars~\cite{Selsing16}. 
We then update it to $\alpha_\lambda = -2.23 \pm 0.01$ based on the second iteration of the \textsf{QSOFitMORE} fitting, using the spectrum with the initial host galaxy model subtracted.
The scale factor of the continuum model needs some consideration.
We first scaled the single power-law so that the emission-line plus power-law quasar model matches the F356W quasar photometry ($22.51\pm0.05$ mag).  
However, when we did so, the equivalent widths of the Balmer absorption lines became too large to be reproduced with the stellar population models.
One could avoid this issue by assuming a flatter $\alpha_\lambda$; however, when we did so, the extracted host continuum became too blue to be reproduced by galaxy SED models.
This is also the case when the assumed \sersic\  index is changed to $n=1,2,$ or $4$.
We therefore allocate additional flux to the quasar component in order to derive meaningful host SED characteristics.
After some experimentation, 
we found that adding $+0.2$ mag to the F356W quasar photometry resulted in a satisfactory result.
This inconsistency between the imaging and spectroscopic decomposition is potentially due to a significant contribution of unresolved galaxy emission to the quasar photometry.
Given the potential oversubtraction of the quasar emission, the stellar population analysis of \targettwo\ warrants caution.
Multiband or higher-resolution imaging follow-up observations are required.

\medskip
\noindent {\bf Strength of Balmer absorption and Balmer break} \label{sec:balmerbreak}

Here we report model-independent measures of the strength of Balmer absorption lines and the Balmer breaks.
To characterize the strength of the Balmer line absorption, we measure the H$\delta$ rest-frame equivalent width for both the original spectrum and the decomposed host galaxy-only spectrum.
Following the procedure of ref~\cite{Goto03},  we measure the inverse-variance weighted mean continuum flux at two wavelength windows: [4030, 4082] \AA, and [4122, 4170] \AA\ and interpolate them to estimate the local continuum at the  wavelength of H$\delta$.  
Using the decomposed host galaxy spectrum, we derive EW(H$\delta$) $=9.5 \pm 0.3$~\AA\ for \targetone, and $=9.1 \pm 0.1 $~\AA\ for \targettwo, which meet the  definition of quiescent galaxies adopted in ref~\cite{WuPF25} ($> 4$~\AA).
The H$\delta$ EW of the original spectrum before subtracting the quasar continuum is $2.8 \pm 0.1 $~\AA\ and $3.9 \pm 0.1 $~\AA, respectively.

We also measure the strength of the Balmer break for \targetone\ using the decomposed photometry.
The F250M and F300M medium-band filters cover the rest-frame wavelengths of 3259--3506~\AA\ and 3825--4266~\AA, thus straddling the Balmer break.
The ratio of the flux density at these two filters  can serve as a proxy for  the Balmer/4000~\AA\ break indices in the literature.
We take the ratio in $F_\lambda$ to derive $F_\lambda(\mathrm{F300M}) / F_\lambda(\mathrm{F250M})=1.9 \pm 0.8$.
The corresponding ratio in $F_\nu$ is $2.8\pm1.2$.
According to ref~\cite{Curtis-Lake23}, this value is comparable to those expected for a single stellar population with age of several 100 Myr  (see their Extended Figure~1), broadly consistent with  the age estimate of our SED fitting analysis (Table~\ref{tab:Table_E4}).
Note that the original wavelength windows applied in ref~\cite{Curtis-Lake23} are [3145, 3563]~\AA\ and [3751, 4198]~\AA.
Also note that there may be contribution of [Ne~III] $\lambda3867$ to the F300M photometry.
We do not measure the same index for \targettwo, as the photometric data is limited to F115W and F356W.

Recently, ref~\cite{Inayoshi_Maiolino24} suggested that an AGN can mimic the stellar Balmer break and Balmer absorption when the nuclear radiation is absorbed by dense gas clouds.
This scenario may explain the continuum shape, and the apparently large stellar masses and high abundance of the dust-obscured compact ``little red dot'' (LRD) galaxies in the early universe.
We argue that such a model is not a good fit  to the two quasars in this paper, which show similar features.
The photometric SEDs in Figure~\ref{fig:spec} are based on the spatially extended components of the two sources.
Therefore, the clear Balmer break in \targetone\ does not originate from AGN emission, which is expected to be unresolved.
The constraint on the Balmer break for \targettwo\ is rather weak, and this object shows a double-peak profile in H$\alpha$ emission (see a later Method section); 
however, the broad Balmer absorption of the high-redshift LRD galaxies is mostly blueshifted with respect to the narrow emission lines~\cite{Inayoshi_Maiolino24}, while  \targettwo's H$\alpha$ has a higher peak at the bluer side than its redder side.
Given these differences, the arguments of ref~\cite{Inayoshi_Maiolino24} would not be applied to \targetone\ and \targettwo, at least not to the same extent as to LRDs.

\medskip
\noindent {\bf Modeling the stellar populations of quasar host galaxies with \textsf{Bagpipes}} \label{sec:sedfit}

We perform spectrophotometric fitting for the two quasar host galaxies using \textsf{Bagpipes}, a public spectral fitting code to model galaxy SEDs  \citep{Carnall18, Carnall19}.
We limit the spectral fitting window to rest-frame $\leq 4,800$~\AA, because there are no apparent absorption lines from the host galaxy at longer wavelengths.
We also find that the use of the full NIRSpec data overly biases the SED fitting toward the rest-frame optical wavelengths, as the relative weight of the NIRCam photometry, particularly F150W that is sensitive to the rest-frame UV wavelengths, becomes reduced, and that the absorption line fitting is not optimal.
The resulting NIRSpec wavelength range for \targetone\ and \targettwo\ is $\lambda_\mathrm{rest} = 3890$ -- $4800$~\AA, and $4010$ -- $4800$~\AA, respectively.
We also mask the red wing of the H$\gamma$ absorption line of  \targettwo\ over $\lambda_\mathrm{rest} = 4344$ -- $4369$~\AA, because this range cannot be well reproduced by our analyses.
A potential reason  for this poor fit is that \targettwo\ exhibits a distinct line profile (Fig.~\ref{fig:J1512_Ha}) that is not represented by Gaussian profiles.
The default stellar population synthesis models of \textsf{Bagpipes} are used, which are based on the 2016 version of the BC03~\cite{BC03, Chevallard16} models with the Kroupa initial mass function~\cite{Kroupa01}.
The same function is commonly adopted in the analyses of massive quiescent galaxies in the literature.
The lower and upper cutoff masses are $0.1~M_\odot$ and $100~M_\odot$, respectively. 
These models use the MILES stellar spectral library~\cite{MILES}, which has a wavelength resolution of $\sim2.5$~\AA\ at $3525$ -- $7500$~\AA.

We explore the SFH of the target galaxies assuming two models: delayed-$\tau$ SFH (SFR$(t) \propto t e^{-t/\tau}$), where $\tau$ represents the decay timescale, and the non-parametric continuity SFH model of ref~\cite{Leja19}.
The choice of the delayed-$\tau$ SFH is appropriate for the target galaxies, because we aim to characterize the SFHs of galaxies that are expected to decline after a major star-forming epoch.
The non-parametric SFH model is more flexible than the parametric SFH, while the continuity prior implemented works against a rapidly quenched SFH.
During the SED fitting process, the stellar age is limited to a range of 10--840 Myr (\targetone) and  10--800 Myr (\targettwo) using a logarithmic prior.
These upper limits correspond to the age of the Universe at the quasar redshifts.
The stellar mass is allowed to vary within the range $\log{(M_*/M_\odot)} =$ 5--15 using a logarithmic prior.
Dust attenuation is modeled with the Calzetti law~\cite{Calzetti00}.
A logarithmic prior is used for the rest-frame $V$-band attenuation over the range $A_V=0.01$--$5$ mag.
Stellar metallicity is fixed to $0.5\ Z_\odot$, a value expected from extrapolation of the stellar mass - gas-phase metallicity relation of $z=4$--$10$ galaxies~\cite{Nakajima23}.
Redshift has a Gaussian prior, given by the measured [O~{\sc iii}] redshift and its estimated uncertainty.
A noise scaling factor ($\times$ 1--10) is implemented to match the flux uncertainty output by the JWST pipeline to the actual variation of the data relative to the model.
 Note that the recovered scaling factor for the two galaxies in this study (1.4--1.9) is broadly consistent with what ref~\cite{Carnall23} find for their quiescent galaxy, suggesting that the error spectrum produced by the JWST pipeline underestimates the actual signal fluctuation.
Stellar velocity dispersion $\sigma_*$ is measured by convolving the stellar continum models with Gaussian kernels in velocity space.
We allowed $\sigma_*$ to vary within the range $\sigma_*=$ 1--1000~km s$^{-1}$.
The instrument resolution ($\sigma\approx170\ \mathrm{km\ s^{-1}}$) at the wavelengths of the Balmer absorption lines is taken into account by referring to the JWST User Document \footnote{\url{https://jwst-docs.stsci.edu/jwst-near-infrared-spectrograph/nirspec-instrumentation/nirspec-dispersers-and-filters}}.
Logarithmic priors are used for these two parameters.
Finally, nebular emission lines are taken into account for \targetone\ with ionization parameter fixed to $\log{U}=-3$.
This nebular component is meant to provide an upper limit of the H$\alpha$ flux.

We reran our SED fitting for \targetone\ by fixing the stellar metallicity from 0.1 to 1 $Z_\odot$ with the delayed-$\tau$ SFH model,  finding that the choice of the stellar metallicity does not significantly affect the stellar mass estimate.
The  mass-weighted age changes from 240 to 350~Myr (50 percentile), while the recovered SFH satisfies the quiescent threshold for all cases, thus not affecting the main results of this paper.
Note that the goodness of fit does not largely change from the fiducial $0.5\ Z_\odot$ model ($|\Delta \chi_\mathrm{reduced}^2|<0.2$), indicating that a direct estimate of stellar metallicity is challenging with the existing data.

We note that the $\sigma_*$ measurements of the two galaxies  based on Balmer absorption lines should be interpreted with caution.
\targetone\ aligns well with the host's major axis, whereas for \targettwo, the slit is tilted by approximately $60$ degrees relative to the disk-like elongated component of the stellar emission. 
In addition, the kinematics of  A- and F-type stars, which dominate the observed spectra of the two galaxies at rest-frame optical wavelengths, may not accurately represent the kinematics of the older stellar populations that typically dominate the galaxy's total mass. 
To achieve more precise $\sigma_*$ measurements, integral field spectroscopy with JWST or ALMA, as well as observations of non-hydrogen $\sigma_*$ tracers~\cite{GH06} --- such as the Calcium triplet $\lambda\lambda8498$, $8542$, $8662$ --- are necessary.
We also address the systematic difference between \textsf{Bagpipes} and \textsf{pPXF} measurements in the next Method section.

\medskip
\noindent {\bf Comparison of different fitting methods} \label{sec:sedfit_ppxf}

In addition to \textsf{Bagpipes}, we also fit the JWST data of the two galaxies with the Penalized PiXel-Fitting method (\textsf{pPXF})~\cite{Cappellari23}, which is widely used in the characterization of kinematics and stellar population of galaxies.
Here we adopt the \textsf{FSPS} stellar library~\cite{Conroy09, Conroy10} to model the same photometric and spectroscopic data as in the \textsf{Bagpipes} analysis.
To estimate the uncertainties, we performed Monte Carlo resampling of the spectrum using the error vector, incorporating the noise scaling factor inferred from \textsf{Bagpipes}, and repeated the fitting 500 times.
We added the model template broadening ($\sigma_\mathrm{template} = 73$ km s$^{-1}$) in quadrature to recover the intrinsic $\sigma_*$.
Figure~\ref{fig:ppxf} shows the best-fit SED model for each galaxy compared with the \textsf{Bagpipes} results.
The best-fit parameters (redshift, $\sigma_*$, and light-weighted age) are reported in Table~\ref{tab:Table_E4}.
For \targetone, the galaxy SED model that returns from \textsf{pPXF} well agrees with that derived from \textsf{Bagpipes}, confirming the robustness of our measurements.
On the other hand, the best-fit $\sigma_*$ for \targettwo\  ($\sigma_*=120$~km s$^{-1}$) is smaller than that from \textsf{Bagpipes}, and it is consistent with an unresolved line within its $1\sigma$ uncertainty.
A part of the reason for this inconsistency is the limited wavelength coverage of the spectrum and  the fact that the stellar continuum blueward of H$\delta$ is not well reproduced by both codes.
It possibly suggests that there is significant contribution of blueshifted H$\delta$ emission.
We therefore adopt the 2$\sigma$ upper limit of $<190$~km s$^{-1}$ as the \targettwo's $\sigma_*$, which is used in the discussions of the main text and relevant figures, instead of the \textsf{Bagpipes} results.

We also fit the galaxies without fixing the stellar matallicity.
In this case, we find that the metallicity uncertainty for \targetone\ spans from sub-solar to near-solar values, despite its rich photometric dataset, which suggests that the currently available data are not sufficient to constrain stellar metallicity.

\medskip
\noindent {\bf Number density of post-starburst quasar hosts at $z\sim6$}

Given the redshift range ($6.18\leq z \leq 6.40$) and absolute UV magnitude range ($-24.0 \leq M_{\mathrm{UV}}\ \mathrm{[mag]} \leq -21.5$)  of the SHELLQs sample in the JWST program, which is drawn from the parent sample of ref~\cite{Matsuoka18}, the discovery of two post-starburst galaxies hosting quasars corresponds to a number density of $\log{n}\ \mathrm{[/Mpc^{-3}]}=-8.8\pm0.3$, where the error reflects the Poisson uncertainty.
This number density is approximately 3 dex lower than those of massive quiescent galaxies at similar redshifts ($\log{n}\ \mathrm{[/Mpc^{-3}]} = -5.8_{-0.8}^{+0.5}$ at $z=7.3$; ref~\cite{Weibel24}), suggesting that the quasar phase is short relative to the massive galaxies' lifetime.
This difference in number density is even smaller than the ratio of the quasar lifetime to the quenching timescale (1--10 Myr vs a few 100 Myr), as discussed in the main text.
A possible explanation is that our selection of quasar-hosting quiescent/quenching galaxies relies on stellar absorption lines, which requires A- and F-type stars to dominate the stellar population and the stellar continuum to remain detectable over the quasar continuum; consequently, our selection is likely incomplete.
It is also possible that most quiescent galaxies at high redshift are {\em not} quenched by quasar activity.

\medskip
\noindent {\bf Double-peak profile in H$\alpha$ emission of \targettwo}

The NIRSpec data of \targettwo\ exhibits an asymmetric double-peak profile in broad H$\alpha$ emission with the blue component stronger than the red component (Fig.~\ref{fig:J1512_Ha}).
The narrow H$\alpha$+[N~{\sc ii}] multiplet is observed at the wavelengths expected from its [O~{\sc iii}] redshift.
It is likely that the double-peak profile of broad H$\alpha$ originates from nuclear scales, as [O~{\sc iii}] is observed as a single line (although it shows some asymmetry in its wings), and there is no signs of a secondary AGN in the NIRCam image (Fig.~\ref{fig:NIRCam_decomposition}b).

Double-peaked broad Balmer lines in AGN have been shown to arise due to the relativistic Keplerian motion of emitting gas in a geometrically thin and optically thick accretion disk~\cite{Eracleous2009}.
Following the method described in ref~\cite{Ward24}, we use the continuum-subtracted spectrum 
to fit the H$\alpha$+[N~{\sc ii}] multiplet with the ref~\cite{Chen1989} circular disk model describing the H$\alpha$ broad emission line region and a double Gaussian model for the narrow emission lines. 
We find that the double-peaked profile is well-described by a circular accretion disk of inclination angle $\sim$26 degrees from face-on with an emitting region extending from 50 to 3000 gravitational radii, an emissivity vs.\ radius power law index of 1.5, turbulent broadening of 800~km s$^{-1}$ within the disk, and a single spiral arm of amplitude 7.7 (in contrast to the rest of the disk), a width of 40 degrees and a phase of 300 degrees. 
The best-fit disk plus narrow emission line model is shown in Figure~\ref{fig:J1512_Ha}.
The parameters describing the disk are typical of the $z<0.4$ disk-emitting AGN population~\cite{Ward24}.  
Before this discovery, the highest redshift double-peak line emitter known was at $z=1.4$ (ref~\cite{Luo2009}); \targettwo\ at $z = 6.18$ is the new record-holder.  
One of the twelve objects in our sample shows a double peak, a fraction consistent with what has been reported at low redshift (3--30~\%; refs~\cite{Strateva03, Eracleous2009, Ward24}).

We note that we cannot fully rule out the possibility of the presence of two accreting SMBHs in a single galaxy~\cite{Xu09, Smith10} with a separation that is not spatially resolved in our NIRCam images.
Candidate binary quasars have been reported with redshifts as high as $z\sim7$ (refs~\cite{Maiolino24b, Uebler24}).

\medskip
\noindent {\bf Ionized gas outflow}

\begin{figure}[tb!]
\centering
\renewcommand\thefigure{E1}
 \includegraphics[width=0.8\linewidth]{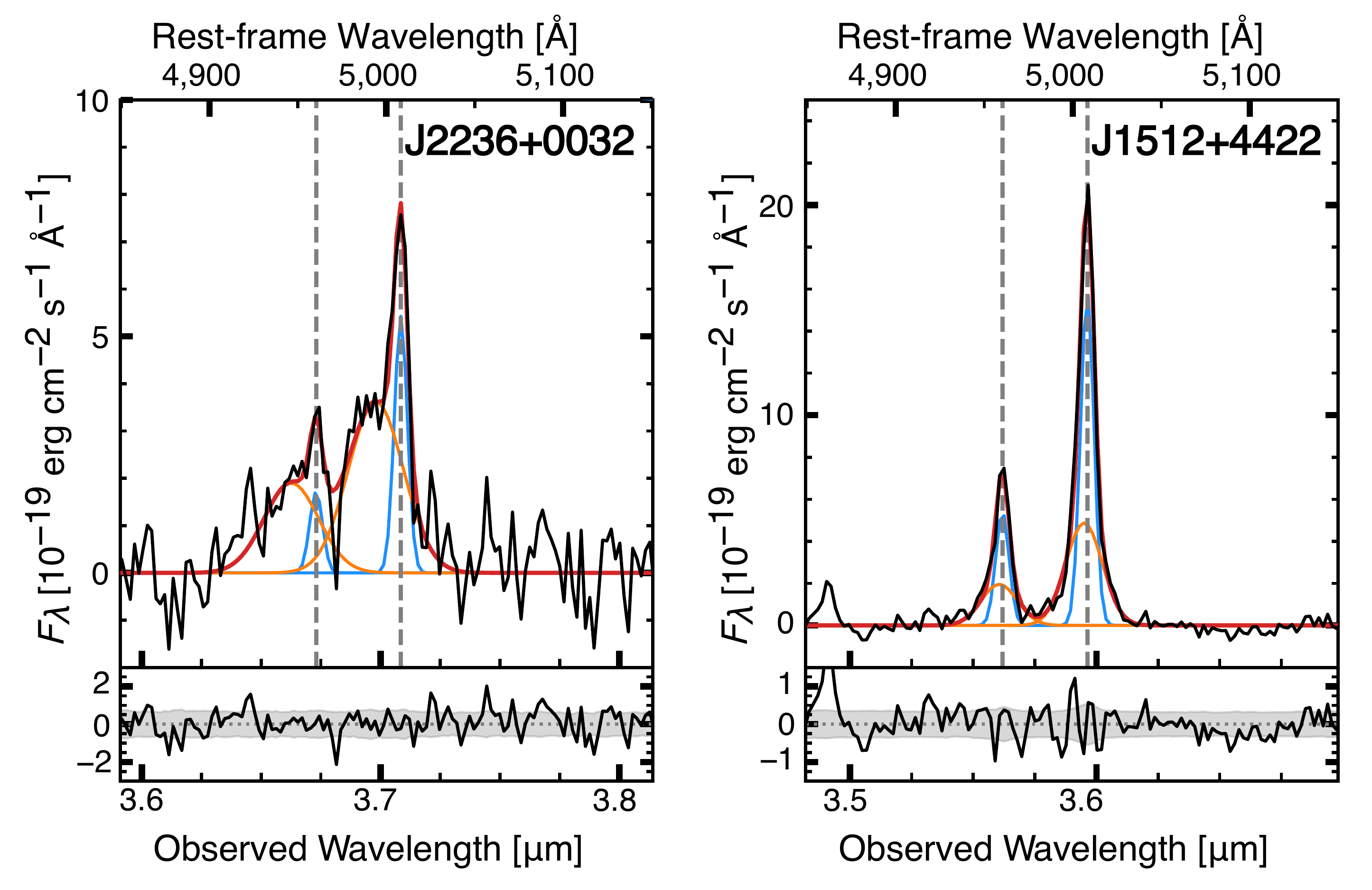}
 \caption{\textbf{[O~{III}] ionized gas profile}.
 Left and right panels show the spectrum of \targetone\ and \targettwo, respectively (black).
 The best-fit models for the continuum, iron, and H$\beta$ line emissions are subtracted for illustration purposes. 
 The core and broad wing components modeled as Gaussian profiles are shown with blue and orange lines, respectively.
 The total profile of the best-fit emission line models are shown with red lines.
 The rest-frame wavelengths  indicated at the top, as well as the expected locations of each doublet line (grey dashed lines) are based on redshifts estimated from the  Balmer absorption lines.
 In the bottom panels, the residuals are shown in black lines, and the $\pm1\sigma$ flux uncertainty at each pixel is indicated as grey shades.
} \label{fig:O3outflow}
\end{figure}

\begin{figure}[p!]
\centering
\renewcommand\thefigure{E2}
 \includegraphics[width=0.8\linewidth]{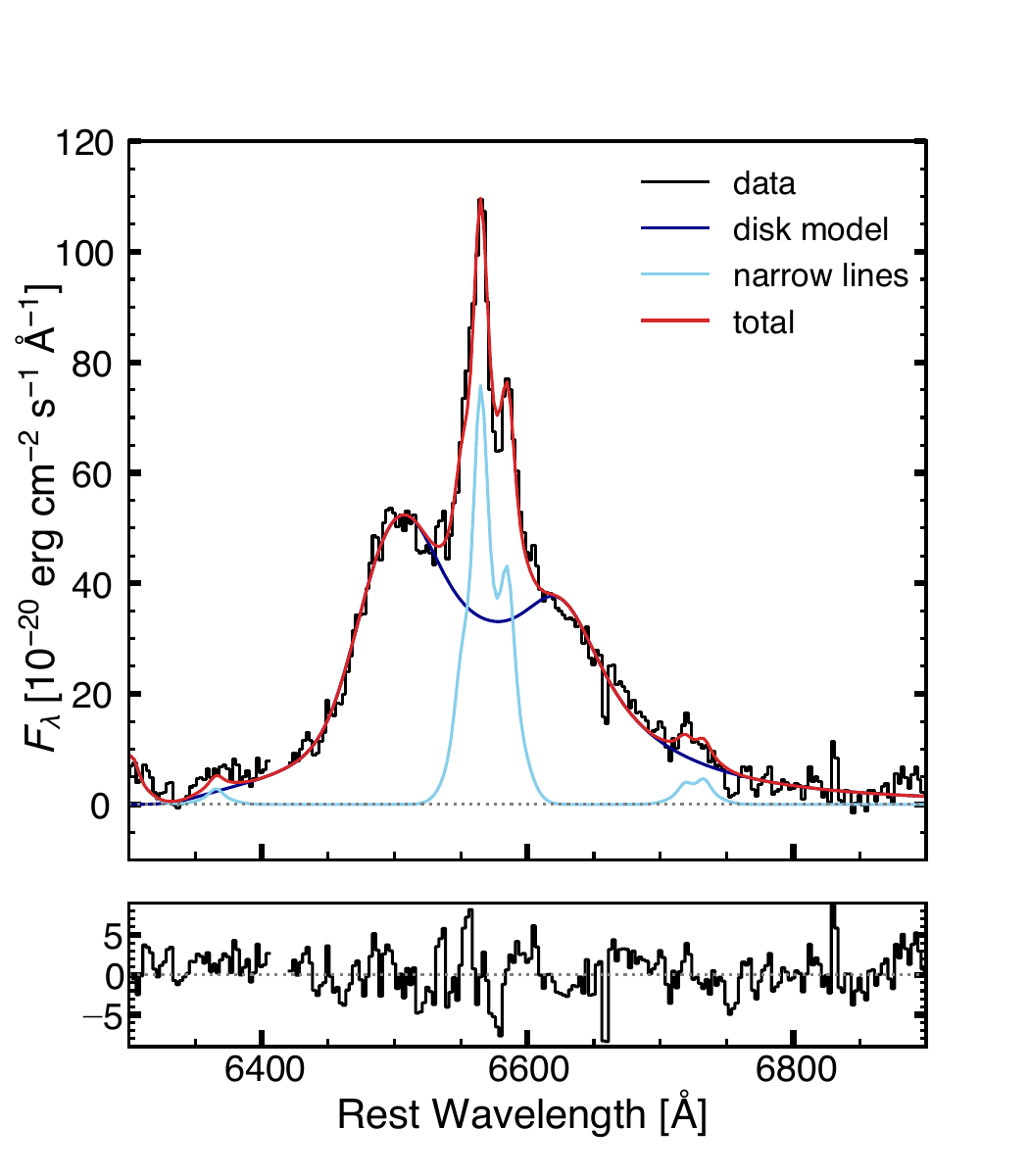}
\caption{\textbf{Double-peaked H$\alpha$ emission for \targettwo.} The observed H$\alpha$ profile (after continuum subtraction) is shown in black. Our model fit based on the method presented in ref~\cite{Ward24} is shown as colored lines (dark blue: disk model, light blue: narrow lines, red: total).
The wavelength and the flux density are presented in the rest frame.
The bottom panel shows the residual fluxes, i.e., the difference between the observed profile and the best-fit model.
} \label{fig:J1512_Ha}
\end{figure}

\begin{figure*}[htb!]
\centering
\renewcommand\thefigure{E3}
 \includegraphics[height=0.4\linewidth]{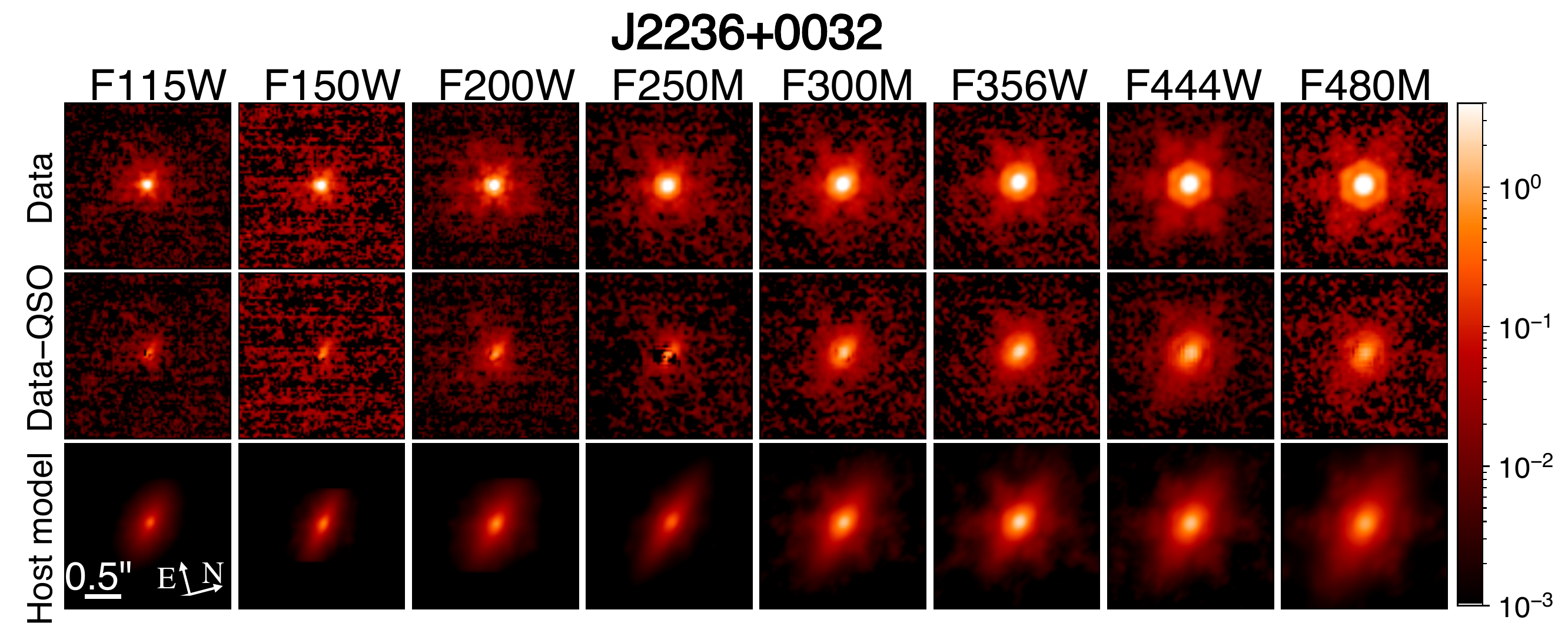}
 \includegraphics[height=0.4\linewidth]{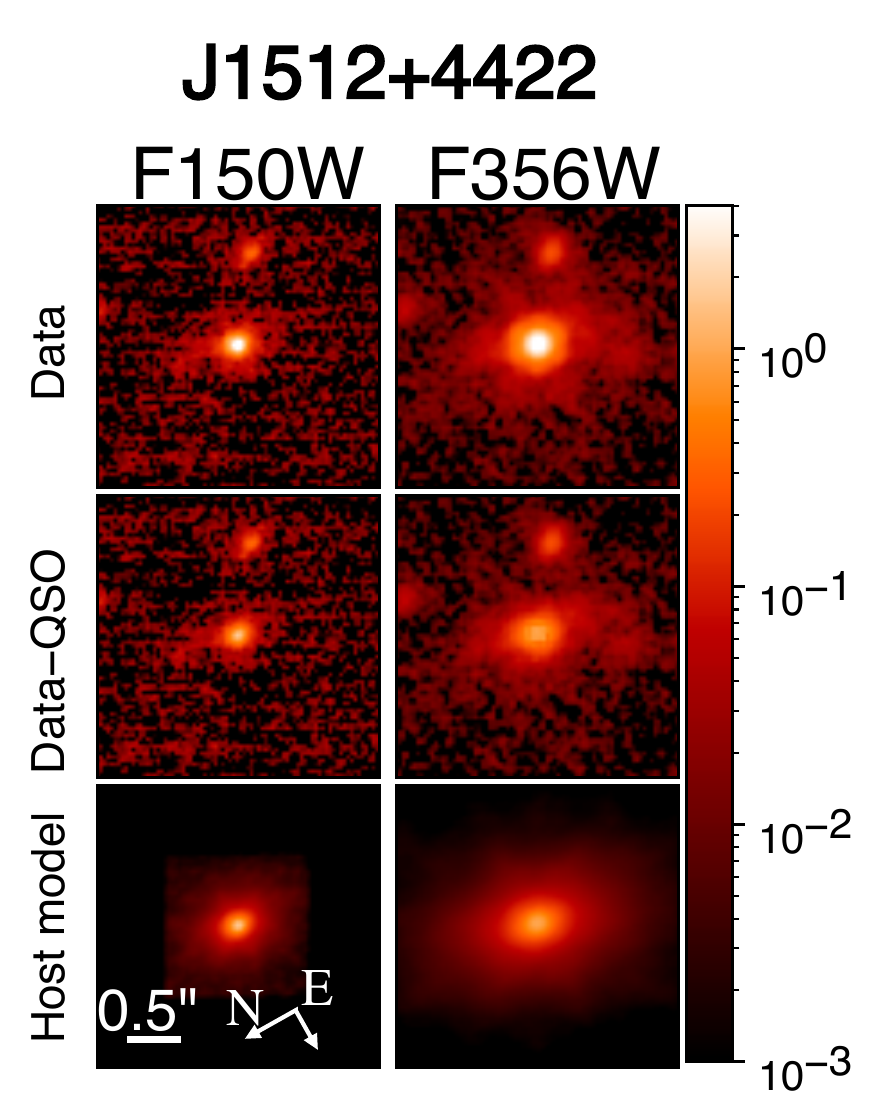}
\caption{
\textbf{Host galaxy detection of \targetone\ (a) and \targettwo\ (b).} Here we present the original NIRCam image (\textit{top}), the PSF-subtracted  \textit{host galaxy-only} image (\textit{middle}), and the best-fit host galaxy model (\textit{bottom}) for each photometric filter from F115W to F480M for \targetone, while F150W and F356W for \targettwo.
The signals are shown in log scale in units of megajansky per steradian.
The image size of each panel is $2\arcsec \times 2\arcsec$.
The arrows in the bottom left panel indicate the North and East directions. 
} \label{fig:NIRCam_decomposition}
\end{figure*}

\begin{figure}[tb!]
\centering
\renewcommand\thefigure{E4}
 \includegraphics[width=0.8\linewidth]{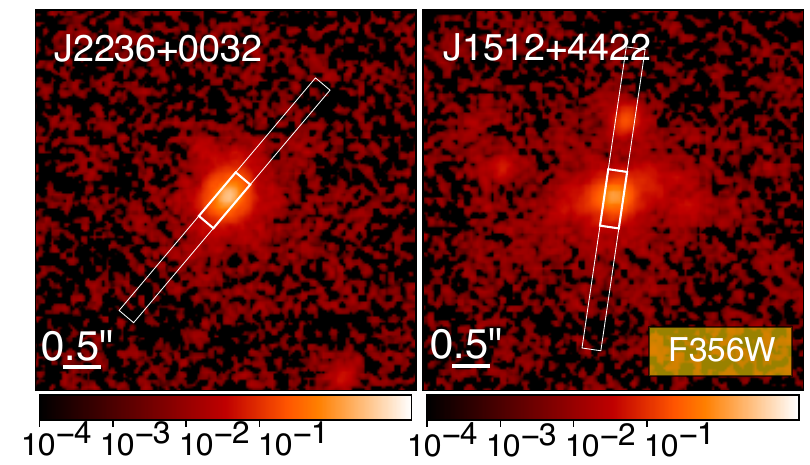}
 \caption{\textbf{NIRSpec Fixed-Slit alignment onto the host galaxies}.
For each galaxy, the outer rectangle indicates the S200A2 slit position and the inner rectangle indicates the extraction aperture of the 1D spectrum.
The background images are the decomposed host images in F356W.
} \label{fig:img_slit}
\end{figure}

\begin{figure}[p!]
\centering
\renewcommand\thefigure{E5}
 \includegraphics[width=0.62\linewidth]{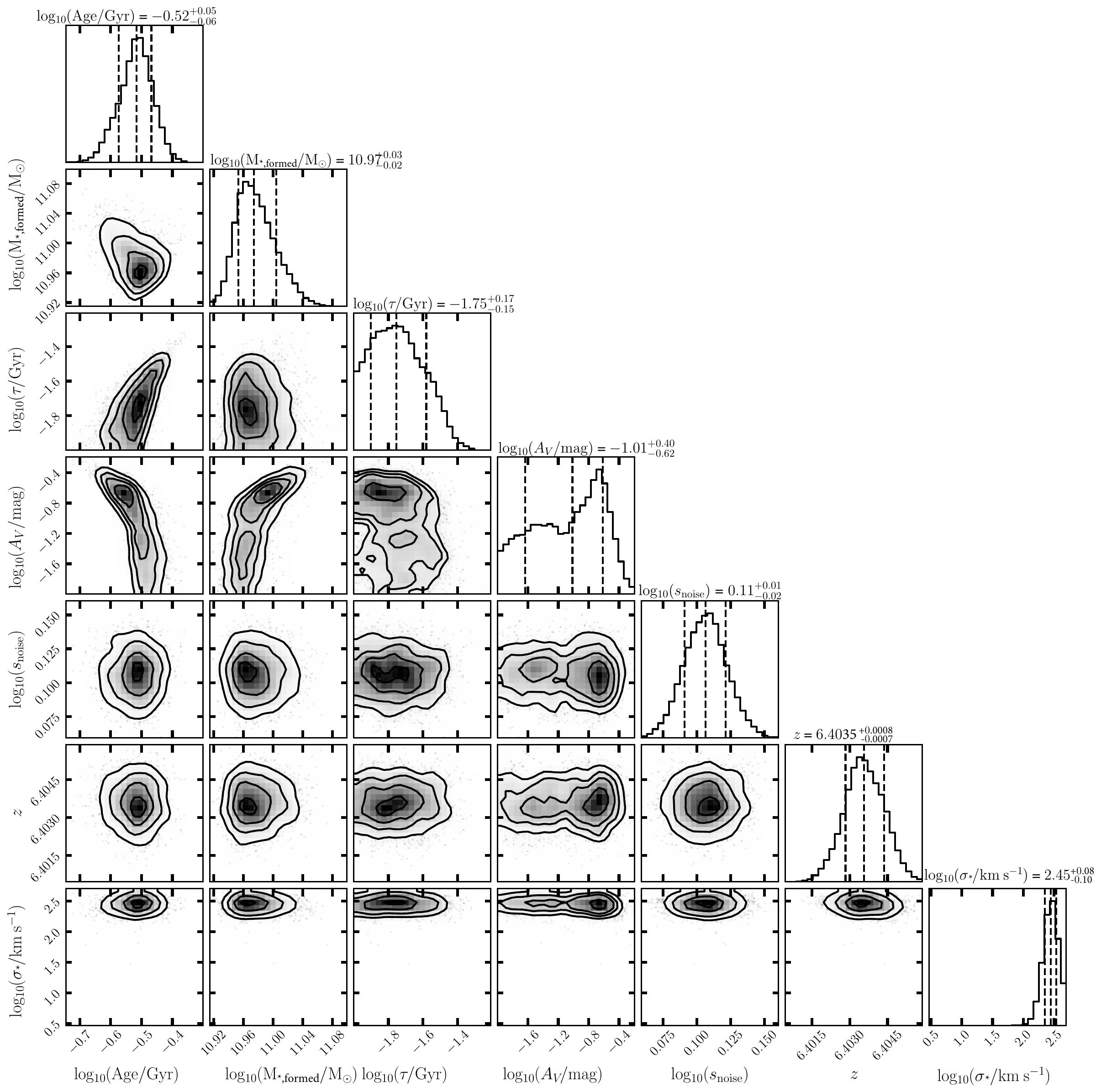}
  \includegraphics[width=0.62\linewidth]{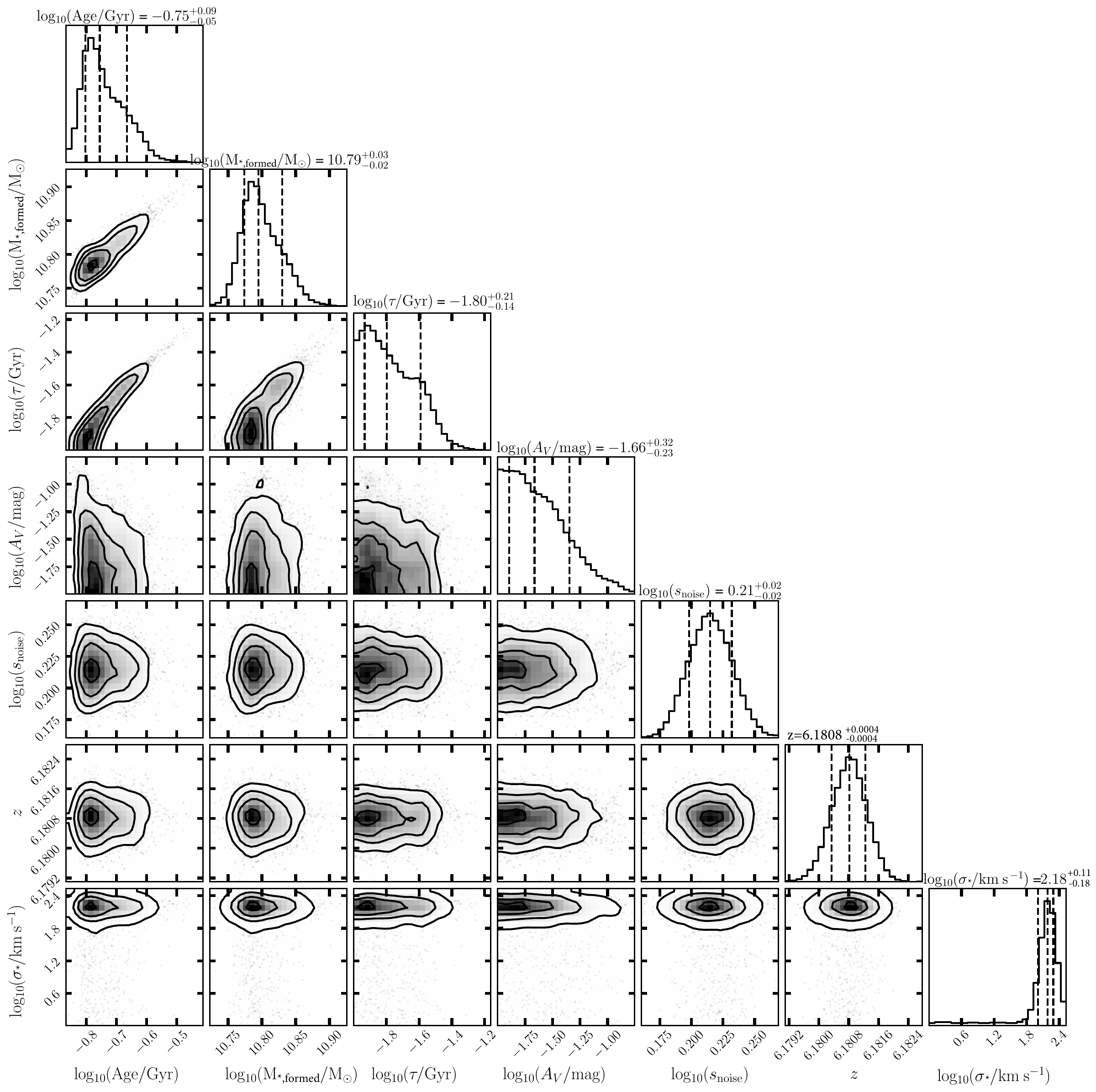}
\caption{\textbf{Example posterior distribution of the Bagpipes SED fitting.}
The top and bottom panels show \targetone, and \targettwo, respectively.
The delayed-$\tau$ SFH model  is presented for each galaxy.
Note that the stellar mass fitted in \textsf{Bagpipes} ($ \log{M_\mathrm{*, formed}}$) represents the total formed stellar mass, from which the observed stellar mass ($\log{M_*}$) is derived.
} \label{fig:corner_J2236}
\end{figure}

\begin{figure}[tb!]
\centering
\renewcommand\thefigure{E6}
 \includegraphics[width=0.8\linewidth]{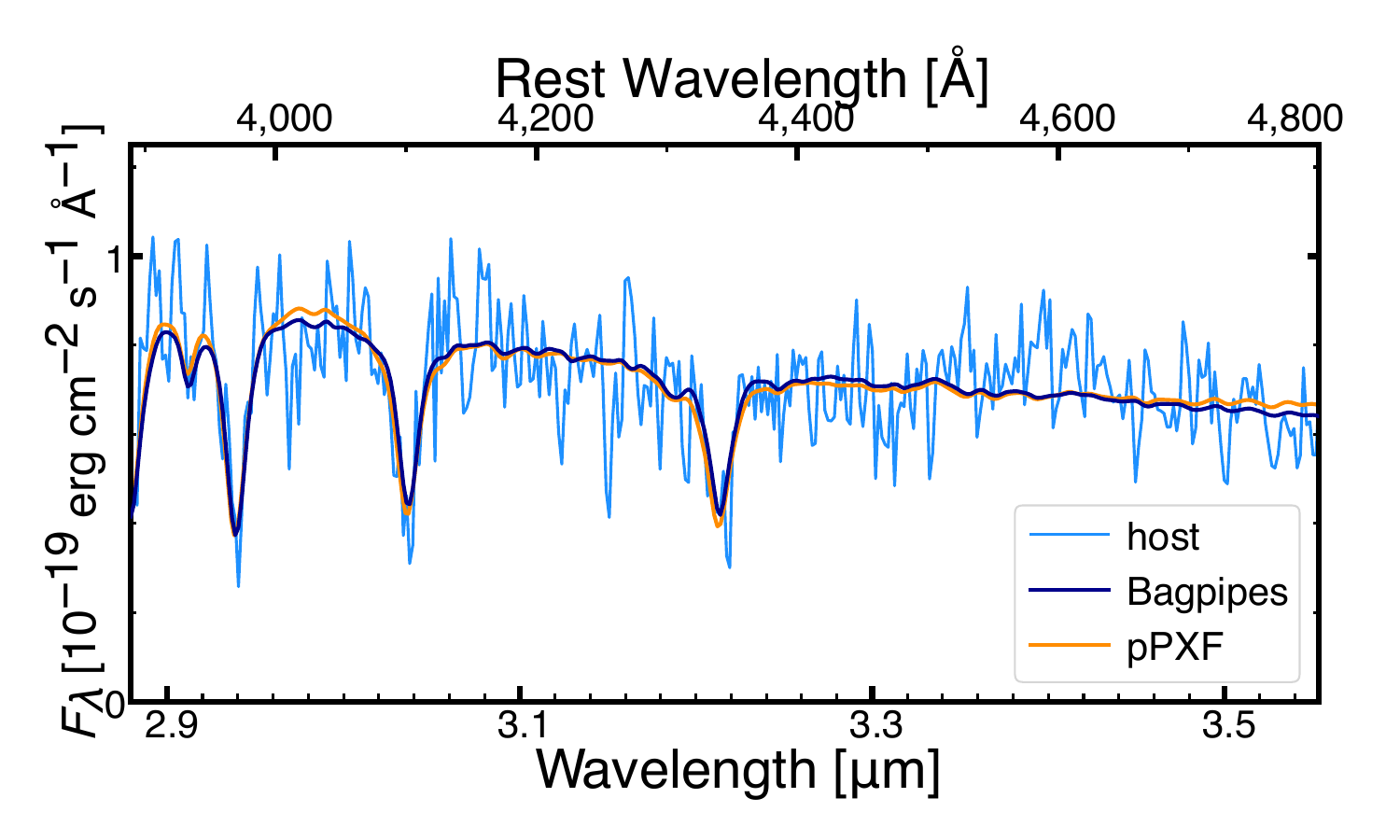}
 \includegraphics[width=0.8\linewidth]{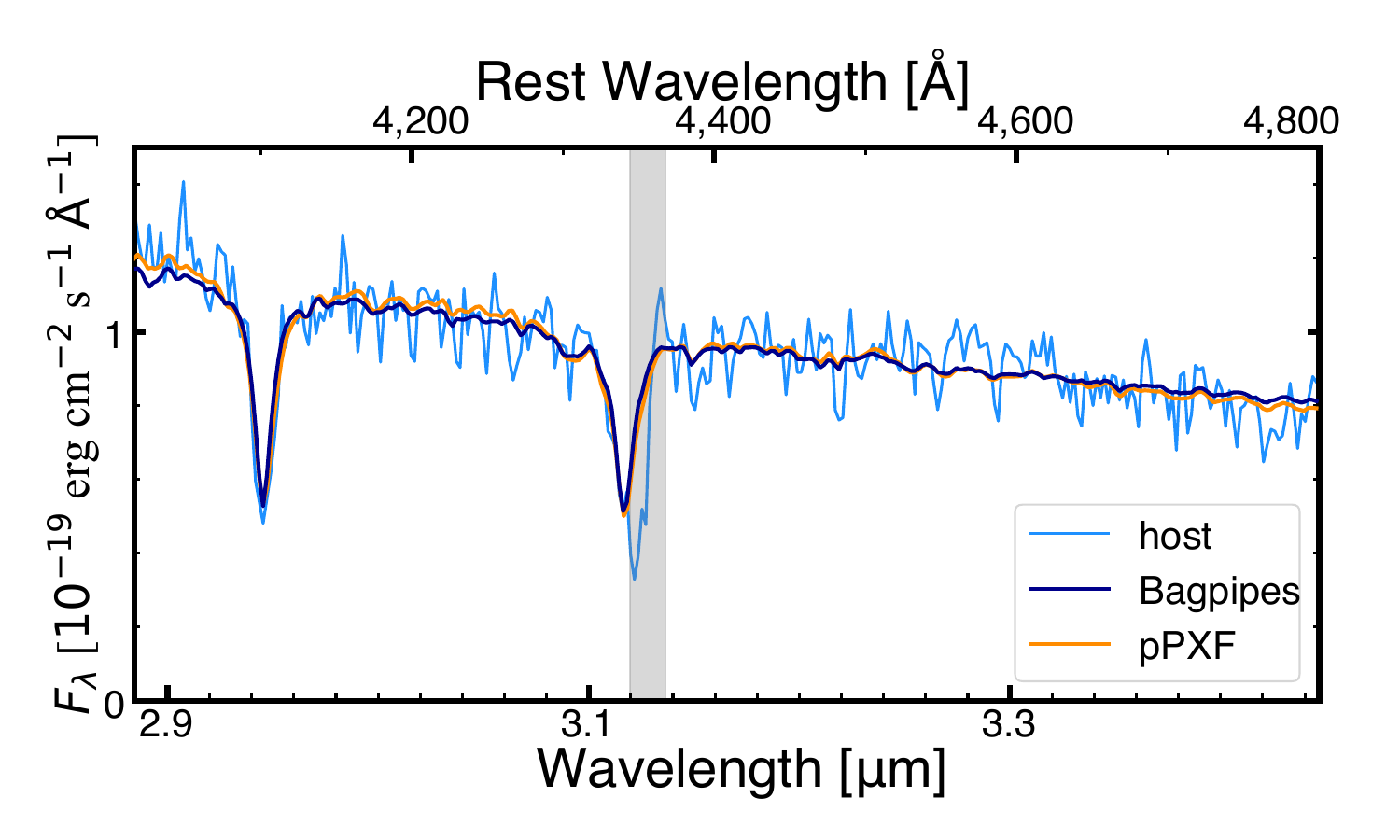}
\caption{\textbf{Comparison of spectral fitting results using two different methods.}
The top and bottom panels show the decomposed host spectrum (light blue) for \targetone\ and \targettwo, respectively. 
The best-fit models from \textsf{Bagpipes} (dark blue) and \textsf{pPXF} (orange) are overlaid.
The red wing of the H$\gamma$ absorption of \targettwo, shown as gray shade, is masked in the spectral fitting.
} \label{fig:ppxf}
\end{figure}

Here, we present the emission line profiles of [O~{III}]~$\lambda\lambda4960,5008$ emission lines for the two galaxies.
Figure~\ref{fig:O3outflow} shows the continuum-subtracted NIRSpec G395M spectra of the two galaxies, with the best-fit Gaussian profiles for H$\beta$ emission lines  also subtracted. 
Using \textsf{QSOFITMORE}, we fit two Gaussian profiles to each doublet line: one representing the narrow ``core'' component and the other representing the broad ``wing'' component, following the method applied in previous studies~\cite{Bischetti17, Fu21}.
We allowed for a velocity shift in the wing component relative to the core component. 
The line profiles of [O~{III}]$\lambda4960$ and  [O~{III}]$\lambda5008$ were fixed to be the same other than the scaling factor.

Table~\ref{tab:Table_E1} summarizes the  [O~{III}] line profile.
We find that both \targetone\ and \targettwo\ exhibit broad wings in [O~{III}] with  FWHM$_\mathrm{broad}=2160\pm140$~km s$^{-1}$ and FWHM$_\mathrm{broad}=1400\pm80$~km s$^{-1}$, while their core components have  FWHM$_\mathrm{core}=330\pm40$~km s$^{-1}$ and FWHM$_\mathrm{core}=350\pm10$~km s$^{-1}$, respectively. 
Such broad components are also seen in other high-redshift quasars~\cite{Marshall23, Yang23}.
Velocity blueshifts are seen for these components with $\Delta v_\mathrm{wing - core}=-850\pm180$~km s$^{-1}$ and  $\Delta v_\mathrm{wing - core} = -110\pm20$~km s$^{-1}$, respectively.
\targetone\ shows the largest velocity offset among the full sample of 12 Subaru HSC quasars observed in our Cycle 1 program.
These wing components account for 50--70\% of the total [O~{III}] flux.
We also find that the  [O~{III}] lines of the two targets are spatially extended across the NIRSpec's S200A2 slit, even beyond the extraction aperture size ($=6$ pixel).
Full analyses of the extended  [O~{III}] lines and their interpretation will be presented in Phillips et al. (in preparation).

\medskip
\noindent {\bf BH mass estimate}

This section presents virial black hole mass measurements based on the broad Balmer emission lines and the decomposed quasar continuum.
For \targetone, we use the prescription of ref~\cite{Vestergaard06} based on H$\beta$.
This single-epoch mass estimate has an intrinsic scatter of 0.4 dex, which is shown in Fig.~\ref{fig:J2236t} and Fig.~\ref{fig:Msigma}. 
The H$\beta$ line width FWHM $=5650\pm160$~km s$^{-1}$ and the 5100~\AA\ continuum luminosity $L_{5100}=(1.78\pm0.01)\times 10^{45}$ erg s$^{-1}$ are measured from the decomposed quasar spectrum.
With a bolometric correction~\cite{Richards06} of $L_\mathrm{bol}=9.26\ L_{5100}$, the inferred bolometric luminosity of \targetone\ is $L_\mathrm{bol}=(1.65\pm0.01)\times 10^{46}$ erg s$^{-1}$.
From these values, we derive an H$\beta$-based BH mass $M_\mathrm{BH}=(1.1\pm0.1) \times 10^9 M_\odot$, and an  Eddington ratio of $L_\mathrm{bol}/L_\mathrm{Edd} = 0.12\pm0.01$.

The BH mass estimate of \targettwo\ is complicated by its double-peaked Balmer lines, which we model as originating from the accretion disk.
We here present the BH mass assuming that the single-epoch method can be applied to double-peak emitters, although this assumption is not fully justified, and thus the BH mass of \targettwo\ is uncertain. 
The FWHM of the broad H$\alpha$ emission line from the model presented in Figure~\ref{fig:J1512_Ha} is $8590$~km s$^{-1}$.
The line luminosity of broad H$\alpha$ is $L_\mathrm{H\alpha}= (4.21\pm0.03)\times10^{43}$ erg s$^{-1}$.
We then derive the H$\alpha$-based BH mass following ref\cite{GH05}, which uses the H$\alpha$ FWHM and $L_\mathrm{H\alpha}$, yielding $M_\mathrm{BH}=1.3 \times 10^9 M_\odot$, and an Eddington ratio of $0.03$.
We use these values in the main text.
We also use the prescription of ref\cite{Vestergaard06}, which is based on H$\beta$ FWHM and monochromatic luminosity at rest-frame 5100~\AA.
The continuum luminosity is measured in the same way as in \targetone, giving $L_{5100}=(1.60\pm0.01)\times 10^{44}$ erg s$^{-1}$.
Using the H$\alpha$ line width as a proxy for H$\beta$ FWHM results in $M_\mathrm{BH}=1.7 \times 10^9 M_\odot$, similar to what we find above. 
The H$\beta$ line profile of \targettwo\ is different from that of H$\alpha$, for which we measure FWHM $=$ 4620~km s$^{-1}$.
Directly applying the recipe of ref\cite{Vestergaard06}, we find $M_\mathrm{BH}=0.48\times 10^9 M_\odot$, and  $L_\mathrm{bol}/L_\mathrm{Edd} = 0.11$.
A more detailed analysis and discussions of the BH mass estimate will be presented in our future papers together with other HSC quasars observed in JWST Cycle 1.\\

\subsection*{Acknowlegements}

We thank Adam C. Carnall for providing supports on the use of \textsf{Bagpipes}.
We thank Yuming Fu for his help on the use of \textsf{QSOFitMORE}.
We thank Jenny Greene, Sune Toft, Takumi Kakimoto and Masayuki Tanaka for fruitful discussions.

This work is based on observations made with the NASA/ESA/CSA James Webb Space Telescope. 
The data were obtained from the Mikulski Archive for Space Telescopes at the Space Telescope Science Institute, which is operated by the Association of Universities for Research in Astronomy, Inc., under NASA contract NAS 5-03127 for JWST. 
These observations are associated with programs GO \#1967 and GO \#3859. 
Support for these programs was provided by NASA through a grant from the Space Telescope Science Institute, which is operated by the Association of Universities for Research in Astronomy, Inc., under NASA contract NAS 5-03127.
This work was supported by World Premier International Research Center Initiative (WPI), MEXT, Japan. 
This work used computing resources at Kavli IPMU.

M.O., X.D., J.D.S., Y.M., T.I., K.I. (Kei Ito), K.K., H.U. are supported by the Japan Society for the Promotion of Science (JSPS) KAKENHI grant Nos. JP24K22894, JP22K14071, JP18H01251, JP22H01262,   JP21H04494,  JP20K14531, JP23K13141, JP17H06130, and JP20H01953.
M.O. and K.I. (Kohei Inayoshi) acknowledge support from the National Natural Science Foundation of China grant Nos. 12150410307 and 12073003, 11721303, 11991052, 11950410493.
K.I. (Kohei Inayoshi) acknowledges support from the China Manned Space Project (CMS-CSST-2021-A04 and CMS-CSST-2021-A06).
S.E.I.B. is funded by the Deutsche Forschungsgemeinschaft (DFG) under Emmy Noether grant number BO 5771/1-1.
Z.H., T.T., M.S. acknowledges support from NSF grant Nos. AST-2006176, AST-1907208, and AST-2006177.
A.L. acknowledges funding from MUR under the grant PRIN "2022935STW".
B.T. acknowledges support from the European Research Council (ERC) under the European Union's Horizon 2020 research and innovation program (grant agreement No. 950533) and from the Israel Science Foundation (grant No. 1849/19).
F.W. (Fabian Walter) acknowledges support from the ERC grant Cosmic\_gas.
J.-T.S. is supported by the Deutsche Forschungsgemeinschaft (DFG, German Research Foundation) under Project No. 518006966.
M.T. acknowledges support from the NWO grant 0.16.VIDI.189.162 (``ODIN'').
S.F. acknowledges support from  NASA through the NASA Hubble Fellowship grant HST-HF2-51505.001-A awarded by the Space Telescope Science Institute, which is operated by the Association of Universities for Research in Astronomy, Incorporated, under NASA contract NAS5-26555.
K.I. (Kazushi Iwasawa) acknowledges support under grant PID2022-136827NB-C44 funded by MCIN/AEI/10.13039/501100011033 /FEDER, EU.
M.V. (Marianne Vestergaard) gratefully acknowledges financial support from the Independent Research Fund Denmark via grant numbers DFF 8021-00130 and 3103-00146.
F.W. (Feige Wang) acknowledges support from NSF award AST-2513040.
R.B. is supported by the SNSF through the Ambizione Grant PZ00P2\_223532.

\subsection*{Author Contributions}
M.O. led the preparation of the observation program, data reduction, spectroscopic data analysis, and manuscript preparation.  
X.D. led the imaging data analysis and contributed to the relevant sections of the manuscript.  
We regard these first two authors as having contributed equally to this work.  
J.D.S. and M.A.S. provided consulting on the manuscript preparation.  
Y.M. contributed to the discovery of the two quasars analyzed in this paper.  
C.W. led the spectral analysis of the double-peak line shape of the H$\alpha$ emission in \targettwo's spectrum.  
C.L.P. performed the NIRSpec 2D spectroscopic analysis and evaluated the strength of the extended H$\alpha$ emission.  
M.T.S. and H.Z. provided theoretical model predictions on the \targetone's SMBH and host stellar growth history presented in Figure~\ref{fig:J2236t}.
K.I. (Kei Ito) contributed to the SED analysis of the two galaxies based on \textsf{pPXF}.
M.O., X.D., J.D.S., Y.M., T.I., M.A.S., C.L.P., and K.J. led the project design and management, also developing the main interpretation of the results.  
I.T.A., K.A., J.A., S.B. R.B., S.E.I.B., A.-C.E., S.F., M.H., Z.H., M.I., K.I., K.I., N.K., T.K., K.K., C.-H.L., J.L., A.L., J.L., T.N., R.O., J.-T.S., M.S., K.S., Y.T., B.T., M.T., T.T., H.U., B.V., M.V., M.V., F.W., F.W., and J.Y contributed to the discussion of the presented results and to the manuscript preparation.

\subsection*{Data Availability}
The JWST data used in this paper (\#GO~1967 \& \#GO~3859) can be accessed via doi:10.17909/mccv-p954.

\subsection*{Code Availability}
The JWST data were processed with the JWST calibration pipeline (\url{https://jwst-pipeline.readthedocs.io}).
Public tools were used for data analyses: \galight\cite{Ding20} and \textsf{QSOFitMORE}\cite{Fu21}.

\subsection*{Competing interests} 
The authors declare no competing interests.

\subsection*{Correspondence} Correspondence and requests for materials should be addressed to M. Onoue (\url{masafusa.onoue@aoni.waseda.jp}) and X. Ding (\url{dingxh@whu.edu.cn}).

\newpage


\renewcommand{\tablename}{Table} 
\setcounter{table}{0} 
\renewcommand\thetable{E\arabic{table}} 
\captionsetup{labelformat=empty}

\begin{table}[p!]
    \centering
    \caption{}\label{tab:Table_E1}
    \begin{tabular}{lcccc}
    \hline \hline
    ID       & $z_\mathrm{[OIII]}$		 & $\Delta v_\mathrm{wing-core}$ & FWHM$_\mathrm{core}$	& FWHM$_\mathrm{wing}$ \\
             &       	   & [km s$^{-1}$] & [km s$^{-1}$] & [km s$^{-1}$] \\
\hline
J2236$+$0032	& $6.4047\pm0.0006$  & $-850\pm50$ & $330\pm40$ & $2160\pm140$ \\
J1512$+$4422	& $6.1806\pm0.0004$  & $-110\pm20$ & $350\pm10$ & $1400\pm80$ \\
\hline
    \end{tabular}
    \flushleft{
    \textbf{Table~E1 $|$ [O~III] line profile.}  
    The [O~III] redshifts are based on the core components.
    $\Delta v_\mathrm{wing - core}$ is the velocity offset of the wing component of  [O~III].    
    The negative values correspond to velocity blueshifts.
    Instrumental broadening is corrected for the line widths.
    }
\end{table}

\begin{table}[p!]
    \centering
    \caption{}\label{tab:Table_E2}
    \begin{tabular}{lcccc}
    \hline \hline
ID &  \multicolumn{2}{c}{\targetone}  &  \multicolumn{2}{c}{\targettwo} \\
\hline
                                & QSO         &    Host        & QSO   &  Host    \\ 
\hline
F115W           & $23.12\pm0.03  $     &   $25.36\pm0.16$       &  $\cdots $       &   $\cdots$ \\
F150W           & $ 22.74\pm0.01  $     &   $24.93\pm0.10$       &  $23.69\pm0.05 $       &   $23.95\pm0.05 $ \\
F200W           & $ 22.26\pm0.01  $     &   $24.44\pm0.09$       &   $\cdots$      &   $\cdots$ \\
F250M           & $ 22.12\pm0.07  $     &   $24.14\pm0.44$       &   $\cdots$      &   $\cdots$ \\
F300M           & $ 22.08\pm0.06   $     &   $23.04\pm0.11$       &   $\cdots$     &   $\cdots$ \\
F356W           & $ 22.04\pm0.10  $     &   $22.87\pm0.20$       &  $22.51\pm0.05 $       &   $22.95\pm0.06$ \\ 
F444W           & $ 21.93\pm0.04  $     &   $22.68\pm0.08$       &   $\cdots$       &    $\cdots$ \\
F480M           & $ 21.42\pm0.01  $     &   $22.85\pm0.05$       &   $\cdots$       &   $\cdots$ \\
\hline
    \end{tabular}
    \flushleft{
    \textbf{Table~E2 $|$ NIRCam photometry of the two quasars and their host galaxies.} 
    The \sersic\ index is fixed to $n=3$ for the host galaxy model in each case.}
\end{table}

\begin{table}
    \centering
    
    \captionsetup{labelformat=empty} 
    \caption{}\label{tab:Table_E3}
    \captionsetup{labelformat=default} 
    \begin{tabular}{cccc}
\hline\hline
ID &  $R_\mathrm{eff}$ &  $R_\mathrm{eff}$ &  $q$  \\
& [$''$] & [kpc] &   \\
\hline
J2236$+$0032 & 0.10$\pm$0.01 & 0.55$\pm$0.06 & 0.34$\pm$0.04 \\ 
J1512$+$4422 & 0.17$\pm$0.02 & 0.96$\pm$0.11 & 0.53$\pm$0.02 \\ 
\hline
    \end{tabular}
    \flushleft{
    \textbf{Table~E3 $|$ Host effective radius $\mathbf{R_\mathrm{eff}}$ and minor-to-major axis ratio $\mathbf{q}$ ($\mathbf{=b/a}$) in F356W.}
    We report the host sizes in both arcseconds and physical scale (proper kiloparsecs).
    }
\end{table}

\begin{table}[htb!]
    \centering
    \caption{}\label{tab:Table_E4}
    \begin{tabular}{lllll}
    \hline \hline
ID &  \multicolumn{2}{c}{\targetone}  &  \multicolumn{2}{c}{\targettwo} \\
\hline
  SFH                             & delayed $\tau$           &    non-parametric        & delayed $\tau$             &  non-parametric     \\ 
\hline
 (Bagpipes)   & & & & \\
$z$                            & $6.4035^{+0.0008}_{-0.0007}$ &  $6.4036^{+0.0007}_{-0.0007}$  & $6.1808^{+0.0004}_{-0.0004}$  &   $6.1808^{+0.0004}_{-0.0003}$        \\ 
$\log{M_*/M_\odot}$            & $10.80^{+0.03}_{-0.02} (\pm0.08) $     &  $10.85^{+0.02}_{-0.02} (\pm0.08)  $     & $10.64^{+0.04}_{-0.01} (\pm0.02)  $   &  $10.68^{+0.04}_{-0.02}  (\pm0.02) $    \\ 
mass-weighted age [Myr]        &  $270^{+20}_{-30}  $        &     $400^{+30}_{-50} $         & $150^{+20}_{-20}  $           &   $190^{+50}_{-30} $     \\ 
$\tau$ [Myr]                   & $ 18^{+9}_{-5}   $         &             $\cdots$                 & $16^{+10}_{-5}   $              &    $\cdots$       \\ 
$A_V$ [mag]                    &$ 0.09^{+0.15}_{-0.07}  $   &  $ 0.05^{+0.08}_{-0.03} $   & $ 0.02^{+0.02}_{-0.01} $     &  $ 0.02^{+0.02}_{-0.01} $      \\ 
$\sigma_*$ [km s$^{-1}$]       &  $290_{-60}^{+50}  $                    &   $280_{-50}^{+50} $   & $160_{-40}^{+30}$        &  $ 150_{-50}^{+40} $        \\
noise scaling factor           & $1.28^{+0.04}_{-0.04}  $     &   $1.28^{+0.04}_{-0.04}$       &  $1.65^{+0.06}_{-0.07} $       &   $1.65^{+0.06}_{-0.05}$ \\
\hline
 (pPXF)   & & & & \\
$z$  &  \multicolumn{2}{c}{$6.3999\pm0.0017$}  &  \multicolumn{2}{c}{$6.1797\pm0.0009$} \\
$\sigma_*$  [km s$^{-1}$]  &  \multicolumn{2}{c}{$270\pm60$}  &  \multicolumn{2}{c}{$<190$ ($2\sigma$)} \\
light-weighted age  [Myr]  &  \multicolumn{2}{c}{$320^{+50}_{-10}$}  &  \multicolumn{2}{c}{$250_{-20}^{+80}$} \\
\hline
    \end{tabular}
    \flushleft{
    \textbf{Table~E4 $|$ Quasar and Host SED parameters obtained from \textsf{Bagpipes} and \textsf{pPXF}}. 
There are two error budgets reported for the stellar mass: one from the SED fitting inference and the other (in parentheses) originating from the decomposed host galaxy photometry in F356W.
Stellar mass uncertainties due to the choice of the IMF are not take into account.
A $2 \sigma$ upper limit on $\sigma_*$ is reported for \targettwo\ from the pPXF measurement  (see Methods).
}
\end{table}

\clearpage

\renewcommand\refname{References (Continued)}

\end{document}